\documentclass[11pt,a4paper]{article}
\usepackage{jheppub}

\usepackage{amsmath}
\usepackage{amssymb}
\usepackage{mathrsfs}
\usepackage{graphicx}
\usepackage{dcolumn}
\usepackage{upgreek}
\usepackage{color}
\usepackage{xcolor}
\usepackage{url}
\usepackage[toc,page]{appendix}
\usepackage[normalem]{ulem}

\usepackage{pdfpages}

\newcommand{\mmiss}{\ensuremath{m_{\textrm{miss}}}}
\newcommand{\mmisssq}{\ensuremath{m^2_{\textrm{miss}}}}
\newcommand{\neu}{{\tilde{\chi}_1^0}}
\newcommand{\signal}{\ensuremath{B^+ \to p \neu}}
\newcommand{\signalSigma}{\ensuremath{B^+ \to \Sigma^+ \neu}}
\newcommand{\signalLambda}{\ensuremath{B^0 \to \Lambda^0 \neu}}
\newcommand{\signalSigmaZero}{\ensuremath{B^0 \to \Sigma^0 \neu}}
\newcommand{\signalLambdaC}{\ensuremath{B^+ \to \Lambda_c^+ \neu}}
\newcommand{\signalSigmaC}{\ensuremath{B^+ \to \Sigma_c^+ \neu}}
\newcommand{\signalSigmaCZero}{\ensuremath{B^0 \to \Sigma_c^0 \neu}}
\newcommand{\signalXiC}{\ensuremath{B^+ \to \Xi_c^+ \neu}}
\newcommand{\signalXiCZero}{\ensuremath{B^0 \to \Xi_c^0 \neu}}
\newcommand{\baryon}{\mathcal{B}}
\newcommand{\signalBaryon}{\ensuremath{B \to \baryon \neu}}
\newcommand{\BtoKnunu}{\ensuremath{B^+\to K^+ \nu \bar\nu}}

\newcommand{\Btopinunu}{\ensuremath{B^+\to \pi^+ \nu \bar\nu}}
\newcommand{\ee}{\ensuremath{e^+e^-}}

\newcommand{\dstl}{\ensuremath{D^{(*)} \ell}}
\newcommand{\tagg}{\mathrm{tag}}

\newcommand{\Br}{\text{Br}}

\newcommand{\eq}{\begin{eqnarray}}
\newcommand{\en}{\end{eqnarray}}
\newcommand{\la}{\langle}
\newcommand{\ra}{\rangle}

\title{Probing $R$-parity violation in $B$-meson decays to a baryon and a light neutralino}

\author[a]{Claudio O. Dib,}
\emailAdd{claudio.dib@usm.cl}
\affiliation[a]{Departmento de F\'isica and CCTVal, Universidad T\'ecnica Federico Santa Mar\'ia,\\ Valpara\'iso 2340000, Chile}

\author[b,c]{Juan Carlos Helo,}
\emailAdd{jchelo@userena.cl}
\affiliation[b]{Departamento de F\'{i}sica, Facultad de Ciencias, Universidad de La Serena,\\
Avenida Cisternas 1200, La Serena, Chile}
\affiliation[c]{Millennium Institute for Subatomic Physics at the High Energy Frontier (SAPHIR), Fern\'andez Concha 700, Santiago, Chile}

\author[d,a,c]{Valery E. Lyubovitskij,}
\emailAdd{valeri.lyubovitskij@uni-tuebingen.de}
\affiliation[d]{Institut f\"ur Theoretische Physik, Universit\"at T\"ubingen, \\
		Kepler Center for Astro and Particle Physics, \\ 
		Auf der Morgenstelle 14, D-72076 T\"ubingen, Germany}
		
\author[f]{Nicol\'as A. Neill,}
\emailAdd{naneill@outlook.com}
\affiliation[f]{Departamento de Ingenier\'ia El\'ectrica-Electr\'onica, Universidad de Tarapac\'a, Arica 1010069, Chile}

\author[g]{Abner Soffer,}
\emailAdd{asoffer@tau.ac.il}
\affiliation[g]{School of Physics and Astronomy, Tel Aviv University, Tel Aviv 69978, Israel}

\author[h,i]{Zeren Simon Wang}
\emailAdd{wzs@mx.nthu.edu.tw}
\affiliation[h]{Department of Physics, National Tsing Hua University, Hsinchu 300, Taiwan}
\affiliation[i]{Center for Theory and Computation, National Tsing Hua University, Hsinchu 300, Taiwan}

\abstract{
We propose a search for $B$ meson decays to a baryon plus missing energy at the Belle~II experiment to probe supersymmetry with a GeV-scale lightest neutralino $\neu$ and $R$-parity violation (RPV).
We perform analytic computations of the signal branching fractions in the framework of effective field theory, with a single nonzero RPV operator $\lambda''_{ij3}\bar{U}_{i}^c\bar{D}_{j}^c\bar{D}_3^c$, where $i,j=1,2$. The hadronic form factors are calculated using an SU(3) phenomenological Lagrangian approach for the proton, as well as several hyperons and charmed baryons.
Since the decay of the neutralino is kinematically and CKM suppressed in this theoretical scenario, it decays outside the detector and appears experimentally only as missing energy.
We detail the analysis techniques at the experimental level and estimate the background in the \signal\ search using published results for $B^+\to K^+ \nu\bar\nu$.
Our final sensitivity plots are shown for both $\lambda''_{113}$  versus the squark mass $m_{\tilde{q}}$ and $\lambda''_{113}/m^2_{\tilde{q}}$ versus the neutralino mass $m_{\neu}$.
We find that the search at Belle~II could probe  $\lambda''_{113}/m^2_{\tilde{q}}$ down to the order of $10^{-8}$ GeV$^{-2}$ in the kinematically allowed $m_{\neu}$ range.
We also obtain current limits on $\lambda''_{123}$ by recasting an existing search interpreted as $\signalLambda$, and comment about searches for $\signalSigma$, $\signalSigmaZero$, $\signalLambdaC$, and $\signalXiC$.
In closing, we briefly discuss potential searches at the LHCb and BESIII experiments.
}
\begin{document}

\maketitle

\section{Introduction}\label{sec:intro}

Supersymmetry (SUSY)~\cite{Nilles:1983ge,Martin:1997ns} has been one of the most sought candidates for physics beyond the Standard Model, in particular because it offers an elegant solution to the hierarchy problem~\cite{Gildener:1976ai,Veltman:1980mj} in the Higgs sector of the Standard Model (SM).
Although no evidence of SUSY has been found so far, the LHC explored wide regions of the SUSY parameter space and has established tight mass bounds on TeV-scale SUSY. 
The minimal supersymmetric extension of the Standard Model (MSSM) in its general form contains lepton-number violating (LNV) and baryon-number violating (BNV) terms, which can induce undesirable proton decay.
One can impose by hand a simple $Z_2$ symmetry, called $R$-parity, which forbids these terms.
Under $R$-parity, all SM particles are even, while all their superpartners are odd.  
If $R$-parity is conserved, not only is the proton stable but so is also the lightest supersymmetric particle (LSP).
Nevertheless, in SUSY with $R$-parity violation (RPV), one can still prevent proton decay by imposing certain conditions
\footnote{See Ref.~\cite{Chamoun:2020aft} for a study on the updated bounds on RPV couplings from nucleon decays.}.
RPV-SUSY models are thus acceptable and, moreover, provide a rich phenomenology that is still only loosely constrained by colliders (see e.g.~Refs.~\cite{Dreiner:1997uz,Barbier:2004ez,Mohapatra:2015fua} for reviews of these models).

The LHC has established strong lower bounds on the masses of heavy SUSY particles, especially color-charged superpartners~\cite{ATLAS:2018nud,CMS:2017brl,CMS:2019vzo,CMS:2019zmd,ATLAS:2020xgt}.
Color-neutral superpartners, however, are not so strictly bounded.
The lightest neutralino $\neu$ can be as light as a GeV or even lighter~\cite{Choudhury:1995pj,Choudhury:1999tn,Belanger:2002nr,Bottino:2002ry,Belanger:2003wb,AlbornozVasquez:2010nkq,Calibbi:2013poa,Gogoladze:2002xp,Dreiner:2009ic,deVries:2015mfw}, in which case its production in decays of $\tau$ lepton or heavy mesons is kinematically allowed.
Such a light neutralino should be dominantly Bino according to current bounds~\cite{Gogoladze:2002xp,Dreiner:2009ic,deVries:2015mfw,Dreiner:2003wh,Domingo:2022emr}.
In addition, it has to decay, e.g., via RPV couplings so as to avoid overclosing the Universe~\cite{Hooper:2002nq,Bottino:2011xv,Belanger:2013pna,Bechtle:2015nua}.
If the RPV couplings are tiny, such light neutralinos can be long-lived so that they appear as displaced signatures or even missing energy in collider experiments.
In this context, in a previous work~\cite{Dey:2020juy} we proposed searching for such a neutralino in $\tau$ lepton decays at Belle~II.
Refs.~\cite{Dercks:2018eua,Dercks:2018wum,Dreiner:2020qbi,Dreiner:2022swd} have estimated the sensitivities of various LHC far detectors, and Ref.~\cite{Candia:2021bsl} investigated the sensitivity of neutrino detectors to atmospheric light neutralinos.

In this work we consider a light neutralino $\neu$ produced in $B$-meson decays through the superpotential RPV terms $\lambda''_{ijk} \bar U_i \bar D_j \bar D_k$, where $U$ and $D$ are the up-type and down-type right-handed supermultiplet fields, $\lambda''_{ijk}$ are dimensionless couplings, and $i, j, k$ are generation indices in the quark-mass basis.
Such interaction terms not only violate $R$-parity but also baryon number.
However, proton decay does not occur in this scenario, as long as no lepton number violating RPV terms are allowed and the neutralino is not lighter than the proton.

We focus on the case in which only one of the RPV couplings $\lambda''_{113}$, $\lambda''_{123}$, $\lambda''_{213}$, or $\lambda''_{223}$ is non-zero and there is no squark mixing.\footnote{In the presence of squark mixing, the experimental results must be interpreted in terms of RPV couplings times mixing parameters. In principle, the different parameters may be disentangled by repeating the measurements proposed here with final states that have different flavor contents, if such decays are kinematically allowed given the neutralino mass.} 
We propose to probe $\lambda''_{113}$ via the decay mode $\signal$ and to study $\lambda''_{123}$ via the decays into strange baryons $\signalLambda$, $\signalSigma$, and $\signalSigmaZero$.
Similarly, the decays into charmed baryons \signalLambdaC, \signalSigmaC, and \signalSigmaCZero\ probe  $\lambda''_{213}$, while \signalXiC\ and \signalXiCZero\ can be used to study $\lambda''_{223}$.
We generically refer to all four modes as $\signalBaryon$, where $\baryon$ indicates one of the above baryons.
These decays can be searched for at an $e^+e^-\to B\bar B$ ``$B$-factory'' facility, namely, the currently running Belle~II experiment~\cite{Belle-II:2010dht,Belle-II:2018jsg} or the BABAR and Belle experiments, which are no longer collecting data. 
The decay of the light neutralino is suppressed, so that it appears as missing energy at the detector level~\cite{Domingo:2022emr}.
While these decays violate baryon number, strictly speaking it is not possible to verify experimentally that the neutralino, which escapes detection, does not carry baryon number.

The superpotential terms $\lambda''_{ijk} \bar U_i \bar D_j \bar D_k$ can also contribute to baryon-antibaron oscillations ($n-\bar n$)~\cite{Calibbi:2016ukt, McKeen:2015cuz} and di-nucleon decay (e.g., $NN\to \pi\pi$, $NN\to KK$)~\cite{Aitken:2017wie}.
However, in the scenario studied here, with no squark mixing, the contributions of $\lambda^{''}_{113}$ and $\lambda^{''}_{123}$ to these processes are suppressed by two weak insertions, yielding limits on $\lambda^{''}_{113}$ and $\lambda^{''}_{123}$ that are much weaker than current LHC bounds and weaker still than the sensitivity of the studies we propose here.
Following Eq.~(23) in Ref.~\cite{Aitken:2017wie}, we can estimate the order of magnitude of the baryon--antibaryon transition amplitude $\delta_{\mathcal B\overline{\mathcal B}}$ 
in terms of the RPV couplings:
\begin{equation}
    \delta_{\mathcal B \overline{\mathcal B}} \sim \frac{\kappa^2 m_{\tilde \chi_1^0}}{m_{\mathcal B}^2-m_{\tilde \chi_1^0}^2}
    \left(\frac{\lambda''_{ijk}g_{1R}^{\tilde q}}{m_{\tilde q}^2}\right)^2,\label{eq:deltaestimate}
\end{equation}
where $\kappa \sim 10^{-2}\,\mbox{GeV}^{3}$ \cite{Buchoff:2015qwa}.
From the LHCb bound  on the $\Xi_b^0-\bar\Xi_b^0$  oscillation rate  \cite{LHCb:2017vth}, namely  $\omega<0.08\,\mbox{ps}^{-1}$, we extract the bound
$\lambda''_{123}/m_{\tilde q}^2 \lesssim 4\times 10^{-4}\,\mbox{GeV}^{-2}$,
where we have taken $m_{\tilde \chi_1^0}=2.5\,\mbox{GeV}$ in Eq.~\eqref{eq:deltaestimate}.
Similarly, the bounds on $\delta_{\mathcal B\overline{\mathcal B}}$ from dinucleon decay listed in Table~I of Ref.~\cite{Aitken:2017wie} can be translated into bounds on the RPV couplings.
The bounds $\delta_{(udb)^2}\lesssim 10^{-13}$, $\delta_{(usb)^2}\lesssim 10^{-10}$,
$\delta_{(cdb)^2}\lesssim 10^{-17}$, and
$\delta_{(csb)^2}\lesssim 10^{-14}$
translate into
$\lambda''_{113}/m_{\tilde q}^2 \lesssim 6\times 10^{-4}\,\mbox{GeV}^{-2}$,
$\lambda''_{123}/m_{\tilde q}^2 \lesssim 2\times 10^{-2}\,\mbox{GeV}^{-2}$,
$\lambda''_{213}/m_{\tilde q}^2 \lesssim 5\times 10^{-6}\,\mbox{GeV}^{-2}$, and
$\lambda''_{223}/m_{\tilde q}^2 \lesssim 2\times 10^{-4}\,\mbox{GeV}^{-2}$,
respectively, again taking $m_{\tilde \chi_1^0}=2.5\,\mbox{GeV}$ in Eq.~\eqref{eq:deltaestimate}.

Decays of a $B$ meson to a baryon and missing energy are motivated also in the context of $B$-mesogenesis models~\cite{Elor:2018twp,Alonso-Alvarez:2019fym,Alonso-Alvarez:2021qfd}.
Signatures of directly detectable or escaping long-lived particles at $B$~factories have also been discussed in Refs.~\cite{Filimonova:2019tuy, Dib:2019tuj, Ferber:2022rsf, Bertholet:2021hjl, Cheung:2021mol,Kim:2019xqj,Kang:2021oes,Acevedo:2021wiq,Dreyer:2021aqd,Duerr:2019dmv,Duerr:2020muu,Chen:2020bok,Dey:2020juy,Bandyopadhyay:2022klg,Guadagnoli:2021fcj}

This paper is organized as follows: section~\ref{sec:prodANDdecay} contains the theoretical framework of the RPV scenario.
This includes the effective Lagrangian with the RPV terms and the calculation of the matrix elements, which we estimate using an effective BNV Lagrangian.
In section~\ref{sec:sensitivity} we explain the experimental analysis technique, estimate the background in the search for \signal, and determine the sensitivity of a Belle~II measurement of the branching fraction $\Br(\signal)$.
We also interpret preliminary BABAR results~\cite{BaBar:2023rer} in terms of \signalLambda\ and compare the sensitivity of this mode to those of \signalSigmaZero\ and \signalSigma.
Furthermore, we discuss the expected sensitivity of decays involving charmed baryons.
Our main physics results are provided in section~\ref{sec:results}, where we extract current limits on $\lambda''_{123}$ and estimate the Belle~II sensitivity to $\lambda''_{113}$.
Finally, section~\ref{sec:conclu} contains our summary, as well as a brief discussion of related measurements that we propose to perform at the LHCb and BESIII experiments. 
We have included an Appendix with details of the form factor calculation for the hadronic transitions.

\section{Theoretical framework: effective Lagrangian and neutralino production and decay}
\label{sec:prodANDdecay}

In this section we describe our theoretical framework. 
We start with an effective Lagrangian needed for the description of 
$\signalBaryon$ decays at the partonic level.
Then we define the effective four-fermion operators 
that generate the matrix elements for the transition hadronic form factors.
Finally, we discuss the computation of the decay rates  $\Gamma(\signalBaryon)$ and present the contributions arising from the different intermediate squarks as a function of the neutralino mass.

The effective Lagrangian $\mathcal{L}_{\text{eff}}$ for the description of $\signalBaryon$ at the partonic level contains two parts: an $R$-parity violating part, $\mathcal{L}_{\text{RPV}}$, with 
trilinear couplings of two quarks and one squark, and 
an $R$-parity conserving (RPC) part describing the couplings of the neutralino $\neu$ with a quark and a squark~\cite{Weinberg:1981wj,Hall:1983id,Faessler:2007br,Li:2007ih}: 
\eq\label{Leff}
\mathcal{L}_{\text{eff}} &=& \mathcal{L}_{\text{RPV}} + \mathcal{L}_{\text{RPC}}
\,, \nonumber\\
\mathcal{L}_{\text{RPV}} &=&  
\sum_{i,j=1}^2\lambda''_{ij3}
\, \epsilon_{abc}   \left(
        \tilde u^{*}_{Ria}\, \bar d_{Rjb}\, b^{C}_{Rc}      
      + \tilde d^{*}_{Rja}\,  \bar u_{Rib}\,  b^{C}_{Rc}
      + \tilde b^{*}_{Ra}\,  \bar u_{Rib}\,  d^{C}_{Rjc} 
        \right) + {\mathrm{h.c.}}\nonumber\\ 
\mathcal{L}_{\text{RPC}} &=& - \sum\limits_{q=u,d,s,c,b} \, g^{\tilde q}_{1R} 
\, \left(\bar q_{R,a} P_L \neu\right) \tilde q_{R,a} + {\mathrm{h.c.}} + \ldots 
\en
where
$a, b, c$ are color indices,    
$\epsilon_{abc}$ is the color three-dimensional Levi-Civita tensor,  
$P_L = (1 - \gamma^5)/2$ is the chiral projector, 
the subscript $R$ indicates a right-handed quark or squark,
and the upper index $C$ denotes charge conjugation. 
Since we take $\neu$ to be a pure Bino, the coupling $g^{\tilde q}_{1R}$ is given by
\eq 
g^{\tilde q}_{1R} = - \sqrt{2} g_W \, e_q \, \tan\theta_W   \,,\label{eqn:bino_coupling}
\en
where $\theta_W$ is the Weinberg angle with $\tan\theta_W \simeq 0.54840$, 
$g_W = e/\sin\theta_W \simeq 0.62977$ 
is the SU(2) weak coupling, and $e_q$ 
is the quark electric charge, i.e. 
$e_u = 2/3$ and $e_d =  e_s = e_b = -1/3$. 
In the last line of Eq.~\eqref{Leff} we explicitly write only the terms with right-chiral fields, since we consider vanishing squark mixing, and only such fields are involved in the RPV interactions considered here.

\begin{figure}[h]
    \centering
    \includegraphics[width=0.32\textwidth]{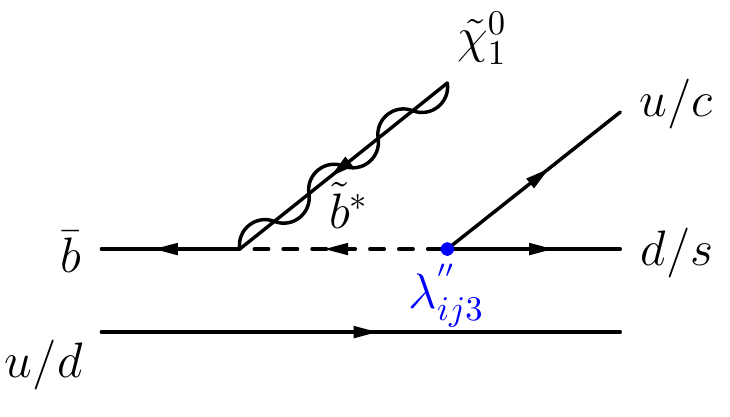}
    \includegraphics[width=0.32\textwidth]{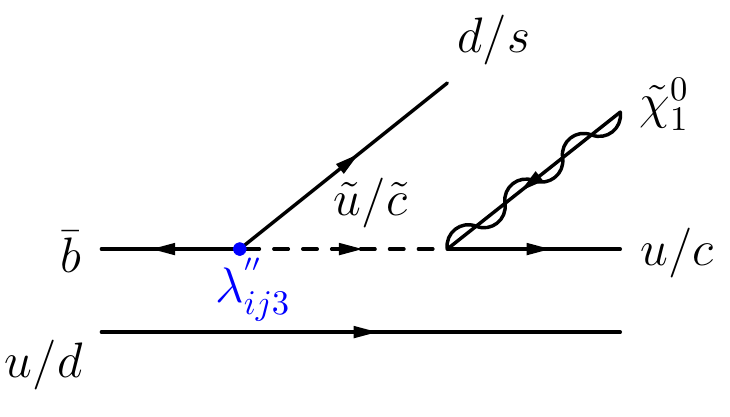}
    \includegraphics[width=0.32\textwidth]{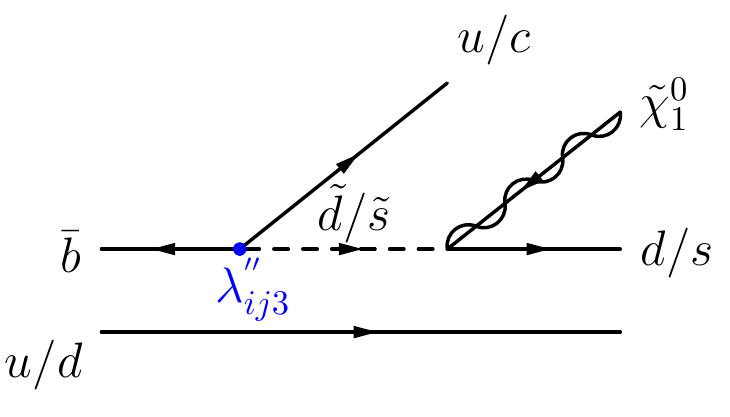}
    \caption{Parton-level diagrams for the decays $\signalBaryon$, where $\baryon$ stands for one of the baryons and $\lambda''_{ij3}$ the corresponding RPV coupling: 
    $p,n$ (for $\lambda''_{113}$); 
    $\Lambda$, $\Sigma^+$, $\Sigma^0$ (for $\lambda''_{123}$); 
    $\Lambda_c^+$, $\Sigma_c^+$, $\Sigma_c^0$ (for $\lambda''_{213}$); 
    and $\Xi_c^+$, $\Xi_c^0$ (for $\lambda''_{223}$).}
   \label{fig:Bdecay}
\end{figure}

The Lagrangian of Eq.~(\ref{Leff}) generates the parton-level diagrams shown in Fig.~\ref{fig:Bdecay}   
for the decays  $\signalBaryon$, as long as the decays are kinematically allowed\footnote{While the decay $B^0\to n \neu$ can also take place, we do not consider it due to the experimental difficulty involved with neutron reconstruction.}. 
To simplify the discussion, we first focus on the \signal\ process, i.e., the diagrams that have a $d$ quark in the final state.

For simplicity, we assume the squark masses to be degenerate and no squark mixing.
In the limit that the squark mass $m_{\tilde q}$ is large, the squark propagators are approximately $1/m_{\tilde q}^2$.
Therefore, the $d$-quark diagrams in Fig.~\ref{fig:Bdecay} correspond to the four-fermion effective Lagrangian 
\eq 
{\mathcal L}_d^{BNV} &=& {\mathcal L}_d^{bud\neu} + {\mathcal L}_d^{udb\neu}  +
{\mathcal L}_d^{dub\neu}  \,, 
\en
where
\eq
{\mathcal L}_d^{q_1q_2q_3\neu} &=& {\mathcal O}_d^{q_1q_2q_3} \,  \neu + {\mathrm h.c.}, 
\en  
the effective operators 
are defined as 
\eq 
{\mathcal O}_d^{q_1q_2q_3} = 
  g_d^{\tilde q_1R} \, {\mathcal O}_{q_1q_2q_3}^{LL} 
\,, 
\en
and we have used the definitions
\eq 
g_d^{\tilde qR} = \frac{g^{\tilde q}_{1R} \lambda''_{113}}{m_{\tilde q}^2}\,, 
\qquad 
{\mathcal O}_{q_1q_2q_3}^{LL} = \varepsilon_{a b c} \,  
\left(\bar q_{3,c} P_L C \bar q_{2,b}\right) \bar q_{1,a} P_L ,
\en
where $C=i \gamma^0 \gamma^2$ is the charge conjugation matrix.

Following Refs.~\cite{JLQCD:1999dld,Aoki:2006ib,Aoki:2008ku,Aoki:2013yxa,Yoo:2021gql}, 
we define the invariant matrix element for the $B^+\to p\neu$ decay as 
\eq 
{\cal M} = \bar u_p(p',s') \, 
\left[W_0^{LL}(q^2) + \frac{\not\! q}{m_{\neu}} W_1^{LL}(q^2)\right] 
\, P_L  \, v_{\neu}(q,s) \,, 
\en 
where $\bar u_p(p',s')$ and $v_{\neu}(q,s)$ are the spinors of
the final-state proton and neutralino, respectively; $p$, $p'$, 
and $q=p-p'$ are the momenta of $B^+$ meson, proton, and neutralino, 
respectively; $m_B$, $m_p$, and $m_{\neu}$ are the masses of these particles;  
and $W_0^{LL}(q^2)$ and $W_1^{LL}(q^2)$ are form factors that parametrize the hadronic 
matrix elements of the BNV three-quark operator  
\eq 
{\mathcal O}^{LL} = g_d^{\tilde bR} {\mathcal O}_{bud}^{LL} 
+ g_d^{\tilde uR} {\mathcal O}_{udb}^{LL} 
+ g_d^{\tilde dR} {\mathcal O}_{dub}^{LL} 
\en 
between the proton and $B^+$ meson states:  
\eq 
\la p| {\mathcal O}^{LL} |B^+\ra = \bar u_p(p',s') \, 
\left[ W_0^{LL}(q^2) + \frac{\not\! q}{m_{\neu}} W_1^{LL}(q^2)\right] 
\, P_L  \,. 
\en 
The form factors $W_0^{LL}(q^2)$ and $W_1^{LL}(q^2)$ can be calculated using  an
SU(3) phenomenological Lagrangian (see Refs.~\cite{Gavela:1988cp} and~\cite{JLQCD:1999dld}-\cite{Yoo:2021gql}) applied to the BNV matrix elements involving a proton and light pseudoscalar mesons $(\pi, K, \eta)$ and extending it to the bottom sector. 
The contribution to the form factors of each of the $\tilde b$, $\tilde u$, and $\tilde d$ squarks is the sum of a
{\it direct} contribution from the $\signal$ process and a
{\it pole} contribution induced by the bottom baryon resonances $\Lambda_b^0$ 
and $\Sigma_b^0$, leading to the virtual two-step process
$B^+ \to p + [\Lambda_b^0 /\Sigma_b^0 \to \neu]$. 

Here, we estimate our matrix elements by extrapolating the strange flavor in 
$\la p| {\mathcal O}^{LL} |K^+\ra$ to the bottom flavor.
In Refs.~\cite{Okubo:1975sc,Liu:2001ce,Dong:2017gaw,Dong:2009tg,Dong:2010gu} detailed discussion 
of possibilities to extend the phenomenological Lagrangians from the SU(3) sector 
to the heavy quark sector have been provided. Derived effective Lagrangians involving 
light and heavy mesons and baryons have been successfully applied to different tasks. 
More concretely, one replaces the decay constant of the Kaon by that of the $B$ meson, $f_K \to f_B$, and 
the masses of the $K^+$ meson and the $\Lambda^0$ and $\Sigma^0$ hyperons by 
the corresponding masses of $B^+$, $\Lambda_b^0$ and $\Sigma_b^0$. 
Our analytical results for the corresponding form factors are shown 
in Appendix~\ref{app:formfactors}.
We also demonstrate in the appendix that the same 
results for the transition form factors can be obtained using an
effective phenomenological BNV Lagrangian constructed in terms of the fields of the $B^+$, 
proton, $\Lambda_b^0$, and $\Sigma_b^0$, which contains some 
adjustable couplings fixed from matching of matrix elements $\la p| {\mathcal O}^{LX} |B^+\ra$ 
in the two approaches. 

The $\signal$ decay width is calculated according to the formula 
\eq\label{decay_width}
\Gamma(\signal) &=& \frac{\lambda^{1/2}(m_B^2,m_p^2,m_{\neu}^2)}{16 \pi m_B^3}
\, \sum\limits_{\text{pol.}} \, |{\mathcal M}|^2 \,, \nonumber\\
\sum\limits_{\text{pol.}} \, |{\mathcal M}|^2 &=&  
    \left(m_B^2 - m_p^2 - m_{\neu}^2\right) \, (\mathcal{A} + \mathcal{B})^2 
- 2 \left(m_B^2 - (m_p+m_{\neu})^2\right) \, \mathcal{A} \cdot \mathcal{B} 
\nonumber\\
&=& \left(m_B^2 - (m_p+m_{\neu})^2\right) \, (\mathcal{A}^2 + \mathcal{B}^2)  
+ 2 m_p m_{\neu}  \, (\mathcal{A} + \mathcal{B})^2  
\,, 
\en 
where 
\eq 
           \mathcal{A} = W_0^{LL}(m_{\neu}^2)  
\,, \qquad \mathcal{B} = W_1^{LL}(m_{\neu}^2)\,,
\en 
and 
\eq 
\lambda(x,y,z) = x^2 + y^2 + z^2 - 2 xy - 2 xz - 2yz 
\en 
is the kinematical triangle K\"allen function. The last line in Eq.~(\ref{decay_width}) shows 
that the matrix element squared $\sum\limits_{pol} \, |{\mathcal M}|^2$ is manifestly positive 
because of the kinematical constraint $m_B > m_p+m_{\neu}$. 

We estimate the rate of the decay $\signalSigma$ in the same way (see details in Appendix~\ref{app:formfactors}). In particular, we replace the $d$ quark by the $s$ quark,
$\lambda''_{113}\to \lambda''_{123}$, $p(duu) \to \Sigma^+(suu)$, and the intermediate baryons $\Lambda_b^0$ and $\Sigma_b^0$ are replaced by the $\Xi_b^0$ and $\Xi_b^{'0}$, which have a $(bus)$ quark-flavor content with an antisymmetric light spin-0 diquark $[us]_0$ and a
symmetric light spin-1 diquark $\{us\}_1$, respectively. 
We also replace the couplings $\alpha$ and $\beta$ that define the matrix elements of the three-quark operators between the proton and vacuum state (see  Eq.~\eqref{eq:3quarkops})  by the respective couplings that define the matrix elements for the case of the $\Sigma^+$ baryon. 
See the discussion in Appendix~\ref{app:formfactors} for more details, as well as for the form-factor calculation for $\signalLambda$ and $\signalSigmaZero$.

\begin{figure}[ht]
    \centering
    \includegraphics[width=0.5\textwidth]{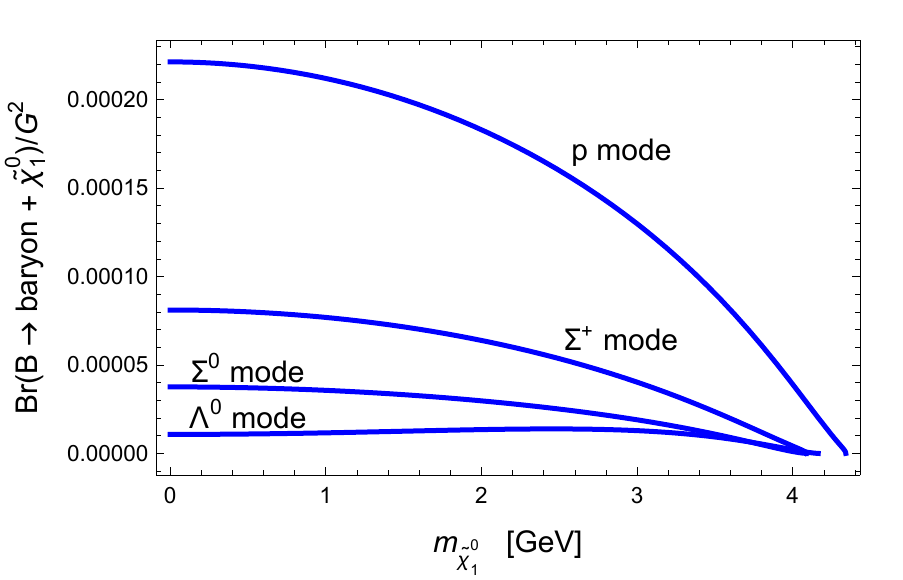}
    \caption{The branching fractions of $\signal$,    $\signalSigma$,    $\signalSigmaZero$, and $\signalLambda$ normalized by $G^2={\lambda''_{1j3}}^2 \times (1 \ {\rm TeV}/{m_{\tilde q}})^4$, as functions of the neutralino mass. 
        \label{Fig2D} 
    }
\end{figure}

\begin{figure}
    \centering
    \includegraphics[width=0.45\textwidth]{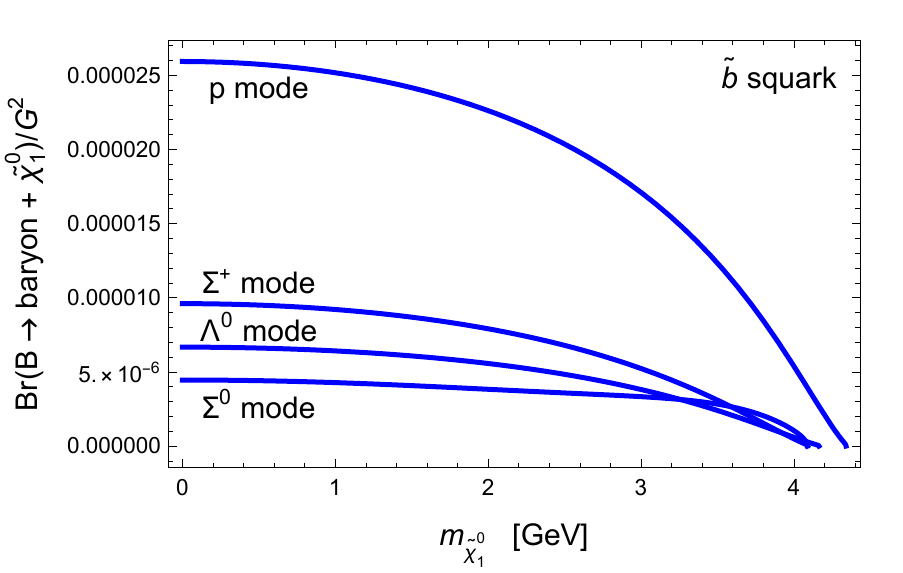}

    \includegraphics[width=0.45\textwidth]{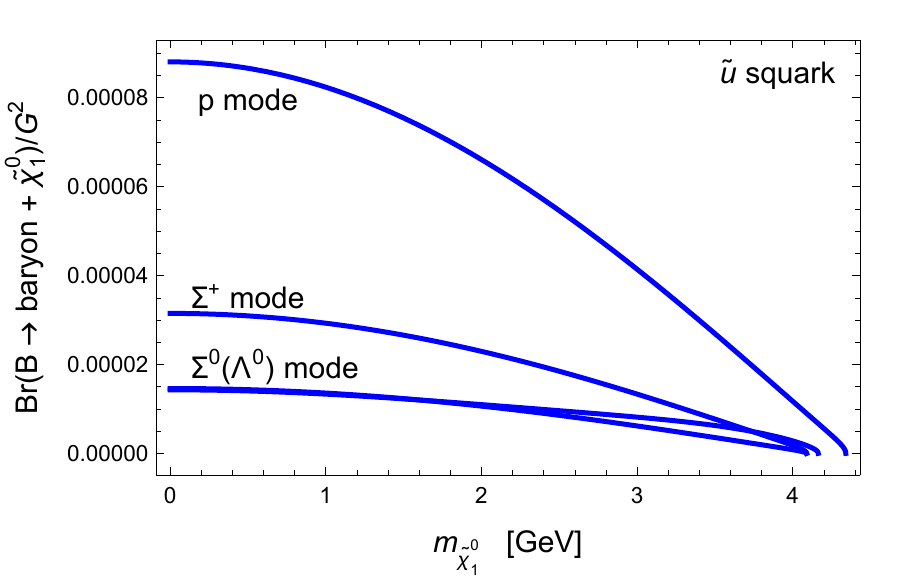}
    
   \includegraphics[width=0.45\textwidth]{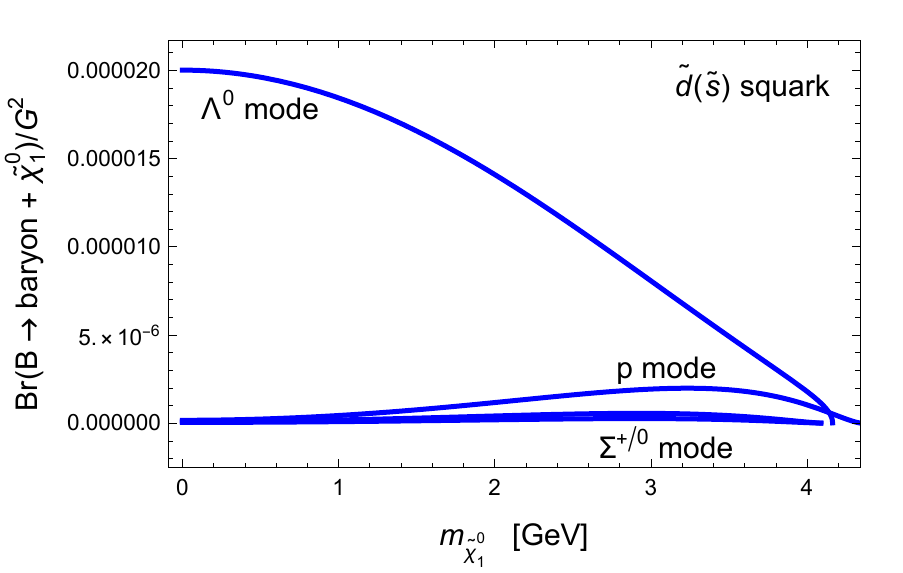} 

   \caption{Individual squark contributions to the branching fractions shown in Fig.~\ref{Fig2D}. } 
        \label{Fig22D} 
\end{figure}

In Fig.~\ref{Fig2D} we compare the four branching fractions  
of interest, normalized by
$G^2={\lambda''_{1j3}}^2 \times (1 \ {\rm TeV}/{m_{\tilde q}})^4$, as functions of $m_{\neu}$.
One can see that $\Br(\signalLambda)$ is significantly suppressed relative to $\Br(\signal)$. This is explained by the compensation of the relative $\tilde b^*$, $\tilde u$, and $\tilde s$ contributions in $\Br(\signalLambda)$, whose magnitudes are shown in Fig.~\ref{Fig22D}.
In particular, at small values of $m_{\neu}$, the ratio $R_1 = \Br(\signalLambda)/\Br(\signal)$ is dominated by the direct contributions to  $W_0^{LL}$ (see Eq.~(\ref{W0LL})), and is given by 
\begin{equation}
    R_1 \simeq \dfrac{3}{8} \, \biggl[ \dfrac{e_u + e_d 
    + 2 e_b}{e_u - e_b} \biggr]^2 = \dfrac{1}{24}.
\end{equation}
In the case of the $\Sigma$ production modes, the corresponding ratio 
$R_2 = \Br(\signalSigmaZero)/\Br(\signalSigma) \simeq 1/2$ is explained 
by the isotopic flavor factor $1/\sqrt{2}$.

Next we extend our analysis to a charmed baryon
produced in the $B$-meson decay. In particular, 
we consider the modes with two possible configuration of two 
light quarks in the charmed baryons $B(cq_1q_2)$, 
where $q_1, q_2 = u, d, s$, --- antisymmetric $[q_1q_2]$ with spin zero 
and symmetric $\{q_1q_2\}$ with spin 1. For detailed classification 
of charmed baryons see, e.g.,  Refs.~\cite{Ivanov:1996fj,Ivanov:1999bk,Faessler:2006ft,Gutsche:2018utw}. 
In particular, in our calculations we consider two states with antisymmetric 
light-quark configurations --- 
$\Lambda^{+}_c=(c[ud])$ and $\Xi^{+}_c=(c[us])$ with masses 
$m_{\Lambda^{+}_c}=2.28646$ GeV and $m_{\Xi^{+}_c}=2.46771$ GeV 
and three states with symmetric light-quark configurations --- 
$\Sigma^{+}_c=(c\{ud\})$, $\Sigma^{0}_c=(c\{dd\})$, and 
$\Xi^{+'}_c=(c\{us\})$ with masses
$m_{\Sigma^{+}_c} = 2.45265$ GeV, 
$m_{\Sigma^{0}_c} = 2.45375$ GeV,  
and $m_{\Xi^{+'}_c} = 2.5782$ GeV. 
In evaluating the hadronic transition form factors, we use the following correspondences to the cases of light baryons: 
(1) for the parton-level diagrams shown in Fig.~\ref{fig:Bdecay} the spectator quarks  
are the same, i.e. $u$ and $d$ for the 
$B^+$ and $B^0$ meson in the initial state, 
respectively; the RPV couplings are 
$\lambda''_{ij3}=\lambda''_{213}$ and 
$\lambda''_{ij3}=\lambda''_{223}$ for 
the ($\Lambda_c^+$, $\Sigma_c^+$, $\Sigma_c^0$) and  ($\Xi_c^+$, $\Xi_c^0$) 
charmed baryon states in the final state, respectively; 
the flavor of other quarks is shown accordingly;  
(2) for the form factors involving $\Lambda^{+}_c$ and $\Xi^{+}_c$ we use the results for the mode with $\Lambda^{0}$; 
(3) for the $\Sigma^{+}_c$ and $\Xi^{+'}_c$ states 
we use the results for $\Sigma^{0}$; and (4) for the $\Sigma^{0}_c$ we use the results for the proton. 
In the evaluation of the pole contributions we include the 
bottom-charmed baryons $\Xi^{+}_{bc} = (b[cu])$ and
$\Omega^{0}_{bc} = (b[cs])$, in which the charm and light quark are in an antisymmetric configurarion, and the baryons
$\Xi^{+'}_{bc} = (b\{cu\})$ and  
$\Omega^{0'}_{bc} = (b\{cs\})$, which have a symmetric configuration. 
The masses involved are  
$m_{\Xi^{+}_{bc}} = 6.933$ GeV, 
$m_{\Xi^{+'}_{bc}} = 6.963$ GeV, 
$m_{\Omega^{0}_{bc}} = 7.088$ GeV, 
$m_{\Omega^{0'}_{bc}} = 7.116$ GeV~\cite{Faessler:2009xn,Branz:2010pq}.  
For the couplings $\beta$ defining the matrix elements of three-quark 
operators between respective charmed baryon and vacuum we use the 
predictions of the QCD sum rules approaches~\cite{Azizi:2008ui,Wang:2009cr,Azizi:2015bxa,Azizi:2015tya}. 
In particular, from results of Refs.~\cite{Azizi:2008ui,Wang:2009cr,Azizi:2015bxa,Azizi:2015tya} we 
deduce the following values of the $\beta$ couplings: 
$\beta_{\Lambda^{+}_c}=  0.835   \times 10^{-2} \ {\rm GeV}^3$,  
$\beta_{\Xi^{+}_c}    =  1.02056 \times 10^{-2} \ {\rm GeV}^3$,  
$\beta_{\Sigma^{+}_c} =  \beta_{\Sigma^{0}_c} = 1.125  
\times 10^{-2} \ {\rm GeV}^3$,  and 
$\beta_{\Xi^{+'}_c}   =  1.375   \times 10^{-2} \ {\rm GeV}^3$.

\begin{figure}
    \centering
    \includegraphics[width=0.45\textwidth]{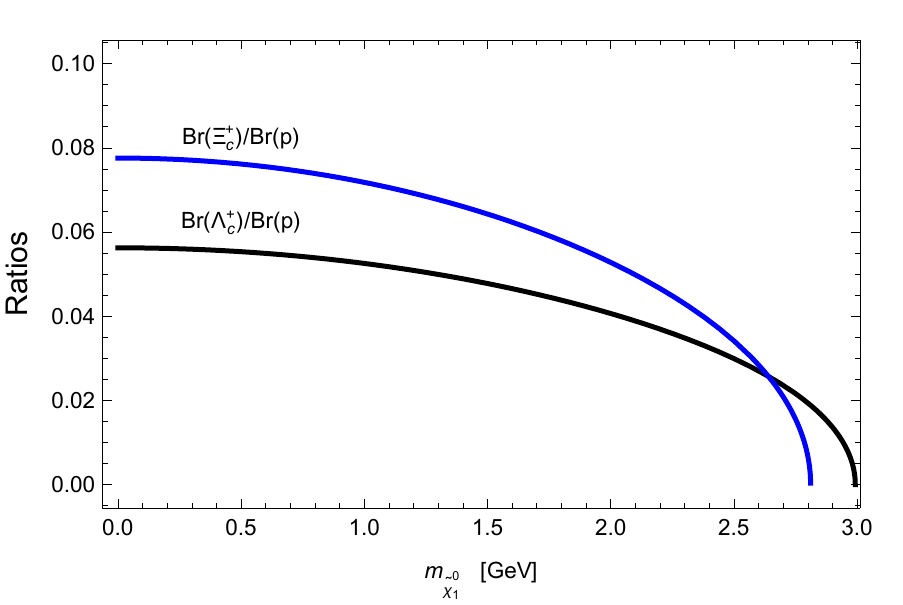}
    \includegraphics[width=0.45\textwidth]{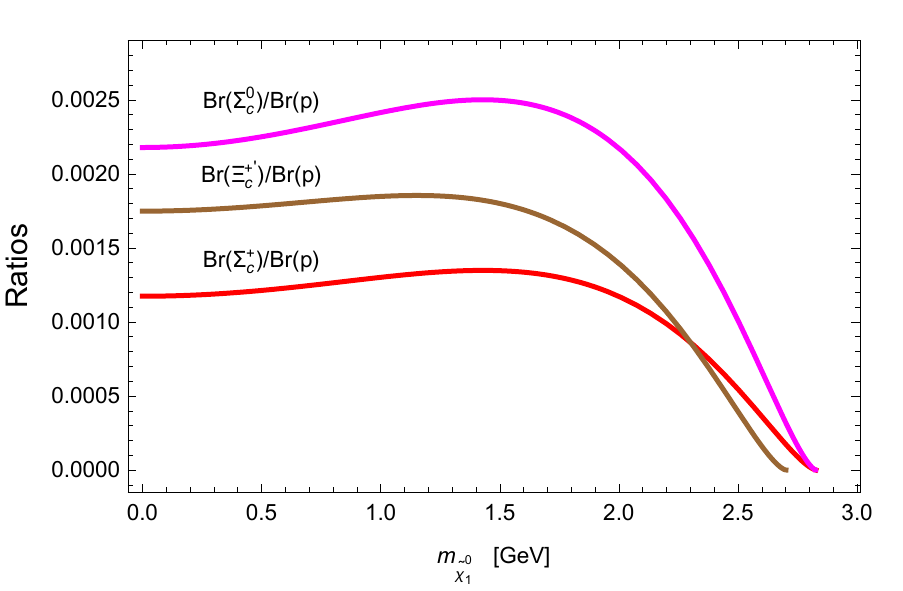}
   \caption{Ratios of branching fractions for \signalBaryon\ decays involving charmed baryons to that of \signal, for (top) the dominant modes with antisymmetric 
   light-quark configurations and (bottom) the suppressed modes with symmetric light-quark configurations.} 
        \label{fig:charmRatios} 
\end{figure}

Our predictions for the ratios of the branchings of the transitions 
of $B^+(B^0)$ into charmed baryons and into the proton 
$\Br(B(cq_1q_2))/\Br(p)$ are shown in 
Figs.~\ref{fig:charmRatios}.
In particular, in upper panel of Fig.~\ref{fig:charmRatios} we display the dominant 
charmed baryons modes with antisymmetric light-quark configuration 
and in lower panel --- the results for charmed baryons 
with symmetric light-quark configurations are shown.

Before moving on to the experimental aspects of the proposed search, we briefly discuss the decay of the lightest neutralino.
For the considered RPV couplings, the neutralino mass range of interest, and the absence of squark  mixing, the neutralino decay is kinematically suppressed by 3 off-shell propagators: a squark, a bottom quark, and a $W$ boson, and is also CKM suppressed.
This renders the lifetime of the lightest neutralino so long that it appears only as missing energy in the detector.

\section{Experimental sensitivity estimation}\label{sec:sensitivity}

The proposed search for $\signal$ is experimentally very similar to those of of $\BtoKnunu$ and $\Btopinunu$, previously performed at the $B$~factories. 
In all these cases, the signal involves a $B$ decay to a single charged particle and unobserved, ``missing'' particles.
We describe the basic analysis techniques in Sec.~\ref{sec:technique} and use previous searches for $\BtoKnunu$ to estimate the sensitivity for \signal\  in Sec.~\ref{sec:bgd}.
In Sec.~\ref{sec:signalLambda} we discuss the relative sensitivities of the modes $\signalLambda$, $\signalSigma$, and $\signalSigmaZero$. 
The sensitivity of the more promising modes involving charmed baryons, \signalLambdaC\ and \signalXiC, is discussed in Sec.~\ref{sec:signalLambdaC}.
In our sensitivity estimates we assume that the search is performed with the full Belle~II data set, which is projected to have an integrated luminosity of $50~\textrm{ab}^{-1}$, corresponding to the production of about $N_{BB}=55\times 10^9$ $B\bar B$ pairs.

\subsection{Analysis technique} \label{sec:technique}

Searches for $\BtoKnunu$ and $\Btopinunu$ have been published by the CLEO~\cite{CLEO:2000yzg}, BABAR~\cite{BaBar:2004xlo, BaBar:2010oqg, BaBar:2013npw},  Belle~\cite{Belle:2007vmd, Belle:2013tnz, Belle:2017oht}, and Belle~II~\cite{Belle-II:2021rof} collaborations.
Using $\ee\to B^+ B^-$ data collected at the $\Upsilon(4S)$ resonance, these searches were performed by reconstructing, or ``tagging'', the decays of one of the $B$ mesons, referred to as the tag $B$.
A single kaon is the only visible particle associated with the signal $B$ candidate, which is assumed to undergo the signal decay of interest.
Tagging is performed via one of three methods: hadronic, semileptonic, or inclusive.
Hadronic tagging has the highest signal purity and lowest efficiency, and the opposite is true for inclusive tagging.
In what follows we briefly describe the three tagging methods for use in the decay \signal.

In hadronic tagging, one attempts to reconstruct the tag $B$ in up to thousands of decay modes involving the Cabibbo-favored quark-level processes $b\to c \bar u d$ and $b\to c \bar c s$. Background is suppressed with selections based on the variables $\Delta E = E_B - \sqrt{s}/2$ and $M_{bc}=\sqrt{s/4 - p_B^2}$, where $E_B$ and $p_B$ are the measured energy and momentum of the tag $B$ in the center-of-mass (CM) frame of the \ee\ collision, and $\sqrt{s}$ is the CM energy. Signal events peak at $\Delta E=0$ and $M_{bc}=m_B$, while background events are more broadly distributed.
Hadronic tagging provides also the squared invariant mass of the unobserved neutralino candidate, the so-called missing mass
\begin{equation}
    \mmisssq = \left(p_{ee} - p_{\tagg} - p_p\right)^2.\label{eq:mmisssq}
\end{equation}
Here $p_{ee}$ is the known 4-momentum of the beam particles~\cite{BaBar:2014omp}, and $p_{\tagg}$ and $p_p$ are the measured 4-momenta of the tag $B$ and of the proton, respectively.
The presence of a signal would appear as a peak in the $\mmiss$ distribution centered at $\mmiss = m_{\neu}$. 

In semileptonic tagging, the tag $B$ is reconstructed in $B^-\to D^{(*)} \ell^- \bar\nu$ decays, where $\ell$ is an electron or muon. 
Because of the unobserved neutrino, the values of $\Delta E$ and $M_{bc}$ are not physically meaningful. 
Rather, one applies the constraint $E_B=\sqrt{s}/2$ to calculating the variable 
\begin{equation}
    \cos\theta_{B-\dstl} = \frac{E_{\dstl}\ \sqrt{s} - M_B^2 - m^2_{\dstl}}{2 p_{\dstl} \sqrt{s/4-M_B^2}},
    \label{eq:mmiss}
\end{equation}
where $E_{\dstl}$, $p_{\dstl}$, and $m_{\dstl}$ are the CM-frame energy, momentum, and invariant mass of the $D^{(*)} \ell^-$ system.
For signal events reconstructed without measurement errors, $\cos\theta_{B-\dstl}$ is the cosine of the angle between the true momenta of the tag $B$ and the \dstl\ system. 
Therefore, signal events are distributed mostly in the range 
\hbox{$-1 < \cos\theta_{B-\dstl} < 1$,} with tails arising from the finite measurement resolution and the spread in $\sqrt{s}$, while background events reach greater negative values.
The unobserved neutrino makes it impossible to determine $p_{\tagg}$.
Nonetheless, the neutralino mass can be calculated with the assumption that the two $B$ mesons are at rest in the CM frame, i.e., $\vec p_{\tagg}=0$ 
in Eq.~(\ref{eq:mmisssq}).
The resulting $\mmiss$ signal peak is significantly broader than for hadronic tagging.

Inclusive tagging is similar to semileptonic tagging in that the tag $B$ decay products are not all observed. 
This may arise from particles that are outside the angular or momentum acceptance of the detector, poor resolution and identification capability for $K_L$ mesons and neutrons, particles with overlapping detector signatures, accelerator background, electronic noise, and limitations of the detector or the reconstruction algorithms.
In Ref.~\cite{Belle-II:2021rof}, the variable used to discriminate between signal and background was the output of a multivariate algorithm, which was trained with simulated events. 
Previous uses of inclusive tagging, beginning with Ref.~\cite{CLEO:1996tag}, used kinematic variables such as $M_{bc}$ for background discrimination.

In all tagging methods, one would require that the event contain only one charged-particle track that is not associated with the tag $B$. 
For \signal, this track should be identified as a proton using the detector's hadron-identification capabilities. 
In addition, the energy deposition $E_{\textrm{extra}}$ in calorimeter clusters that are not associated with the proton or the tag-$B$ decay products must be small in order to suppress background from neutrons and $K_L^0$ mesons. 
As in $\BtoKnunu$ searches, this requirement can be applied directly on $E_{\textrm{extra}}$, or by including this variable as input to a multivariate algorithm. 

In both $\BtoKnunu$ and other studies of rare $B$ decays with final-state neutrinos~\cite{Soffer:2014kxa} the three methods have similar sensitivities for a given data set.
Therefore, all can be used for the search of \signal, yielding similar sensitivities.
In what follows, we use hadronic tagging for our sensitivity estimation.

\subsection{Background and sensitivity estimate for \boldmath{\Br(\signal})}
\label{sec:bgd}

We estimate the \mmiss-dependent background in the  hadronic-tagging analysis from Fig.~8 of Ref.~\cite{BaBar:2013npw}, which shows the \mmiss\ distribution in the BABAR $\BtoKnunu$ search, performed with a data sample of $0.471\times 10^9$ $B\bar B$ pairs.
The background estimate is $0.2$ events per 100-MeV-wide bin for $1<\mmiss<2.5$~GeV, thereby rising linearly to $3.5$ events per bin at $\mmiss=4.2$~GeV.
We scale the background estimate to the Belle~II data set by multiplying it by the ratio $R_L = N_{BB}(\mathrm{Belle~II}) / N_{BB}(\mathrm{BABAR}) = 55\times 10^9 / 0.471 \times 10^9$ of the numbers of $\ee\to B\bar B$ events in the final Belle~II data set and in Ref.~\cite{BaBar:2013npw}, respectively. 

\begin{figure}[h]
    \centering
    \includegraphics[width=0.45\textwidth]{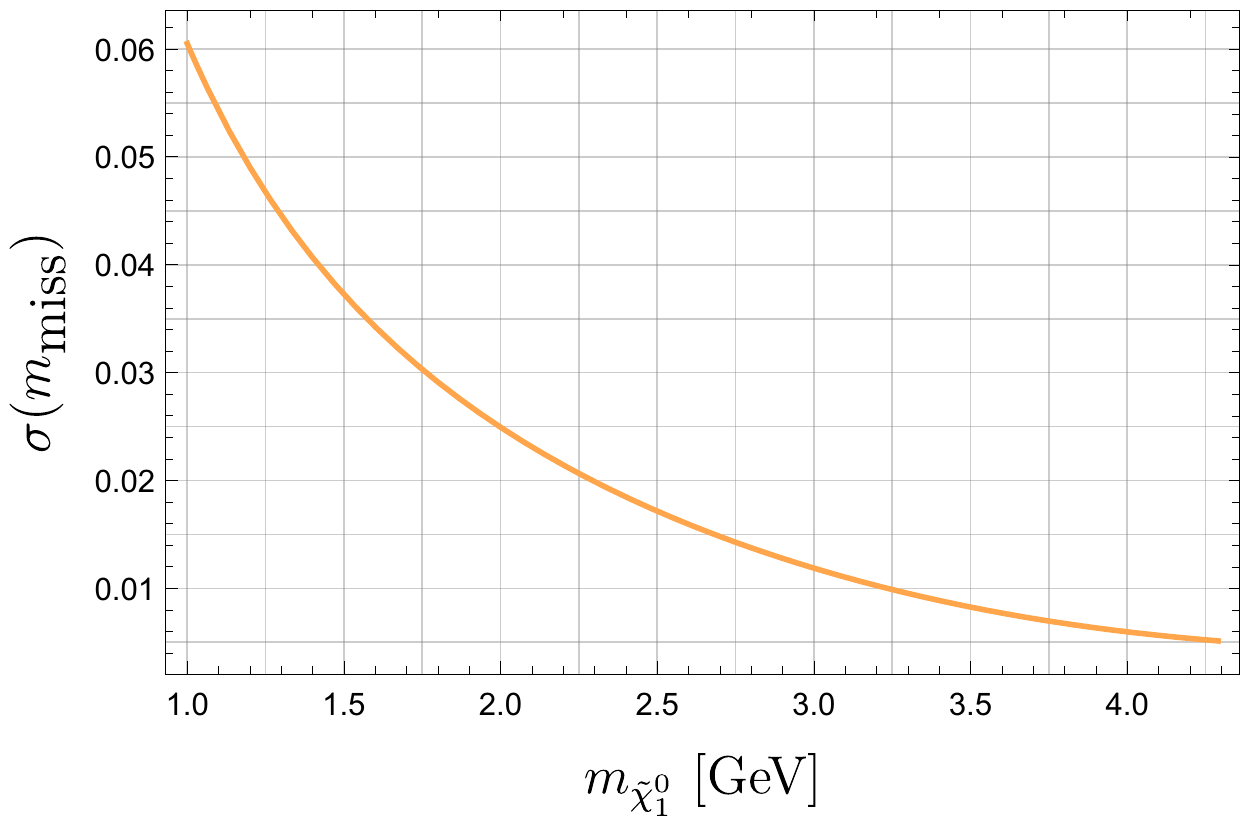}\ \ \
    \includegraphics[width=0.475\textwidth]{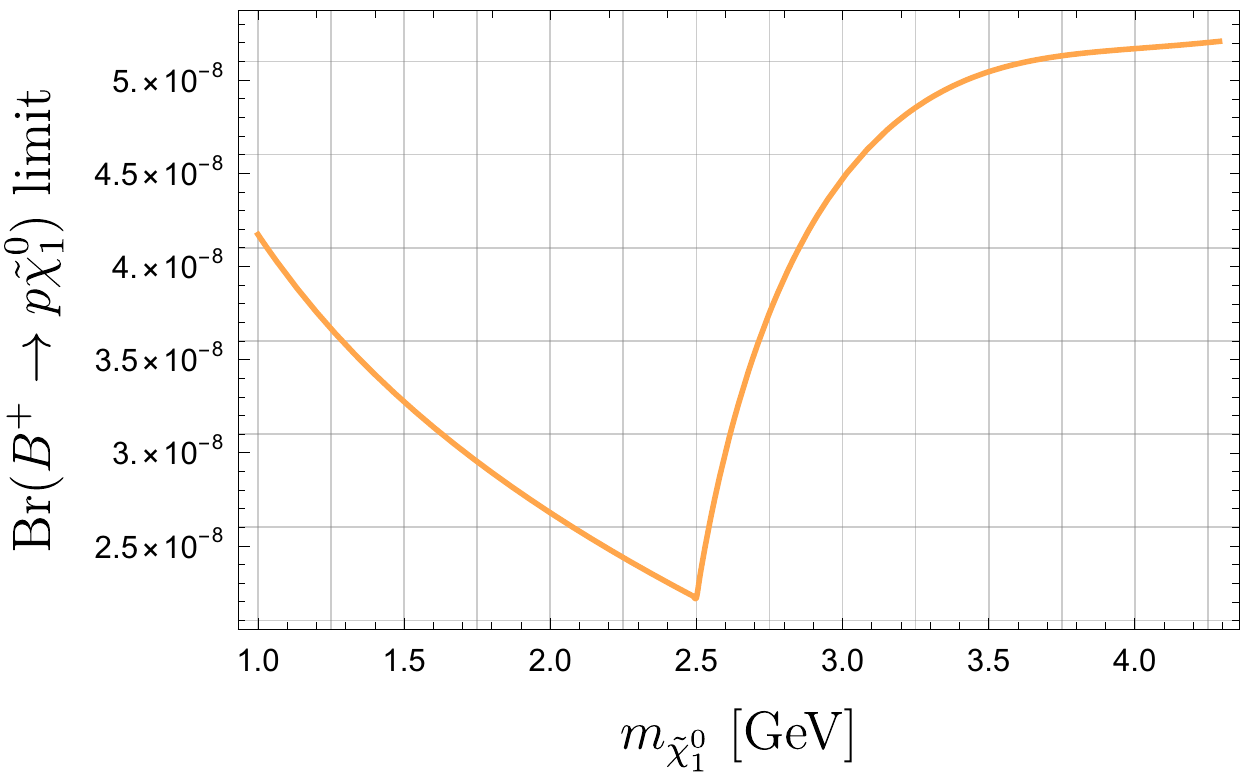}
    \caption{The missing-mass resolution $\sigma(\mmiss)$ (left) and the expected Belle II 90\% confidence-level upper limit on the branching fraction $\Br(\signal)$ as functions of the neutralino mass $m_\neu$.
    The upper limit decreases with the decreasing value of $\sigma(\mmiss)$ but increases for $m_\neu > 2.5~$GeV due to the increasing background. The kink at $2.5~$GeV results from the simple parameterization of the background governed by the limited data available in Ref.~\cite{BaBar:2013npw}.}
   \label{fig:res-BR-limit}
\end{figure}

The estimated background is  reduced relative to that in \BtoKnunu\ by the fact that the proton production rate in $B^+$ decays is about $R_{p/K}\approx 1/16$ of $K^+$ production. 
Specifically, $\Br(B^+\to K^+ +\mathrm{anything})=(66\pm 5)\%$~\cite{Zyla:2020zbs}. While this was measured by ARGUS~\cite{ARGUS:1993vpc} and CLEO~\cite{CLEO:1987ygn} for a nearly equal admixture of $B^+$ and $B^0$, it is not expected to depend significantly on the isospin state of the $B$ meson.
Therefore, we take this branching fraction to apply to the $B^+$.
By contrast, $\Br(B\to \bar p/p +\mathrm{anything})=(8.0\pm 0.4)\%$~\cite{Zyla:2020zbs}, which is about 8 times smaller.
Furthermore, correlating the charge of the proton with the charge of the tag $B$ will reduce the background by a factor of two, given that such a correlation was not performed in the $\Br(B\to \bar p/p +\mathrm{anything})$ measurement.
Baryon production is expected to be similarly suppressed in the non-$B\bar B$ background, specifically $\ee\to q\bar q$, where $q=u,d,s,c$ is a light quark.
Furthermore, this background is reportedly smaller than the $B\bar B$ background in the $\BtoKnunu$ searches~\cite{Belle-II:2021rof, Belle:2017oht, Ferber:2022rsf}.
Thus, multiplication by $R_{p/K}$ gives the expected background yield $N_b^{100}$ for the \signal\ search at Belle~II in each $100$-MeV-wide \mmiss\ bin. 
From this, we calculate the number of background events in a \mmiss\ bin with a width of 2-standard-deviation,
\begin{equation}
    N_b = N_b^{100} \, \frac{2\sigma(\mmiss)}{100~{\rm MeV}}.
\end{equation}
The \mmiss\ resolution $\sigma(\mmiss)$ is dominated by the resolution on the transverse momentum $p_T$ of the proton for most of the \mmiss\ range.
For the Belle experiment, this has been measured to be~\cite{BaBar:2014omp}
\begin{equation}
    \left(\frac{\sigma(p_T)}{p_T}\right)^2 = \left(0.0019  \frac{p_T}{\rm GeV}\right)^2 + \left(0.003 \frac{1}{\beta}\right)^2, 
\end{equation}
where $\beta$ is the charged-particle velocity, which we calculate on average assuming a uniform angular distribution for the protons. 
We ignore the fact that the $p_T$ resolution is expected to be somewhat better for Belle~II, which has a larger drift chamber.
A smaller contribution to $\sigma(\mmiss)$ arises from the  spread of the collider center-of-mass energy, which we take to be 5~MeV. 
The value of $\sigma(\mmiss)$ as a function of $m_\neu$ is shown in Fig.~\ref{fig:res-BR-limit}. 

The 90\% confidence-level (CL) upper limit on the \signal\ branching fraction at Belle~II, assuming the absence of a signal, is calculated from
\begin{equation}
 \Br(\signal) < \frac{1.64 \sqrt{N_b}}{N_{BB}(\mathrm{Belle~II})\;  \epsilon} \, ,
\label{eq:basic-sensitivity}
\end{equation}
where $\epsilon\approx 95\times 10^{-5}$ is the total reconstruction efficiency~\cite{BaBar:2013npw}.\footnote{The efficiency is taken from Table~VI of Ref.~\cite{BaBar:2013npw}, where it is given multiplied by $\Br(B^+\to J/\psi K^+)$.}
We check this branching-fraction sensitivity estimate against an inclusive-tagging study of $B^+ \to K^+ a$, where $a$ is an axion-like particle that is long lived enough to always escape the detector~\cite{Ferber:2022rsf}. Fig.~10 of Ref.~\cite{Ferber:2022rsf} gives the 90\% CL limit Br$(B^+ \to K^+ a) < 10^{-7}$. Scaling this by $\sqrt{R_{p/K}}$ yields Br$(\signal)< 2.5\times 10^{-8}$, in agreement with our hadronic-tagging-based estimate.

\subsection{Sensitivity estimate for \boldmath{\Br(\signalLambda)}, \boldmath{\Br(\signalSigma)}, \boldmath{\Br(\signalSigmaZero)}}
\label{sec:signalLambda}

The Belle collaboration has published a search for $B^0\to \Lambda^0 \psi_D$, where $\psi_D$ is an invisible dark-matter particle~\cite{Belle:2021gmc}, obtaining limits on the branching fraction in $500$~MeV-wide steps of $\mmiss$. 
A preliminary result from BABAR~\cite{BaBar:2023rer} shows tighter limits in the range $1 <\mmiss <4.2$~GeV.  
In Sec.~\ref{sec:results} we reinterpret the BABAR search in terms of $\signalLambda$ and extract limits on $\lambda''_{123}$.

Considering the $\signalSigmaZero$ signal, we note that the dominant decay mode of the $\Sigma^0$ is $\Sigma^0\to \Lambda^0 \gamma$. 
Since the $\Sigma^0 - \Lambda^0$ mass difference is only $77$~MeV, the photon is soft, so its reconstruction incurs lower efficiency and higher background than for $\signalLambda$ reconstruction. 
The efficiency loss may be avoided by  foregoing the photon reconstruction  altogether. 
However, this results in an additional contribution of order the photon energy to the  $\mmiss$ resolution.
Furthermore, while the production rate of $\Sigma$ baryons in $B$-meson decays has not been measured, one can assume that it is similar to that of $\Lambda^0$.
As a result, background from true $\Sigma$ baryons is not smaller than that arising from true $\Lambda$ barons. Overall, although  $\signalSigmaZero$ appears theoretically more promising than $\signalLambda$ in Fig.~\ref{Fig2D}, these experimental difficulties mean that it does not offer an advantage over  $\signalLambda$.

In the search for $\signalSigma$, one must reconstruct the $\Sigma^+$ in the final state $p\pi^0$, which has a branching fraction of 51.6\%~\cite{Zyla:2020zbs}. 
About 48.3\% of the $\Sigma^+$ decays are to $n\pi^+$.
However, since the neutron interacts only hadronically, the reconstruction of this mode is much less practical. 
The efficiency for reconstruction of the soft $\pi^0$ is in the low tens of percent~\cite{SDey}. 
This approximately offsets the advantage of the $\signalSigma$ mode seen in Fig.~\ref{Fig2D}.

\subsection{Sensitivity estimate for \boldmath{\Br(\signalLambdaC)} and \boldmath{\Br(\signalXiC)}}
\label{sec:signalLambdaC}

As seen in Fig.~\ref{fig:charmRatios}, among the modes involving charmed and charmed-strange baryons, \signalLambdaC\ and \signalXiC\ are advantageous due to the large form factors.
Hence we focus on these only. 
Since experimental information involving similar modes is lacking, our sensitivity estimates in this case are crude.

The most favorable final state for $\Lambda_c^+$ decay is $pK^-\pi^+$, with a branching fraction of about 6.3\%~\cite{Zyla:2020zbs}.
We take the reconstruction efficiency for this mode to be 70\% the efficiency of \signal, due to the additional tracks and particle-identification requirements.

To estimate the background, we note that a $\Lambda_c^+$ or $\bar\Lambda_c^-$ is produced in $3.6\%$ of $B$-meson decays~\cite{Zyla:2020zbs}. 
This arises mostly from the Cabibbo-favored quark-level transition $\bar b\to \bar c u \bar d$, which leads to $\bar \Lambda_c^-$ production from a $B^0$ or $B^+$ decay.
By contrast, our signal involves $\Lambda_c^+$ production, which  can proceed via the Cabibbo-favored decay $\bar b\to \bar c c \bar s$.
In this case, the decays with the lightest final-state hadron combination are $B^+\to \Lambda_c^+ \bar p^- \bar D_s^+$ and $B^0\to \Lambda_c^+ \bar n D_s^-$ for which the sum of hadron masses is $5.193$~GeV and $5.194$~GeV, respectively.
Since the $B^+$ and $B^0$ masses are only $5.279$ and $5.280$~GeV, respectively, these decays are highly phase-space suppressed.
The other alternative for $\Lambda_c^+$ production in $B$ decays is $\bar b \to \bar u c \bar s$, which is suppressed by $V_{ud}/V_{cb} \approx 0.08$~\cite{HFLAV:2022pwe}.
Thus, we expect that $\Br(B\to \Lambda_c^+ + {\textrm{anything}}) \lesssim 0.4\%$ which is 20 times less than $\Br(B^+\to p + {\textrm{anything}})$.
The background yield is further reduced by $\Br(\Lambda_c^+\to pK^-\pi^+)$ and the $70\%$ efficiency hit, thus being 450 times smaller than in the search for $\signal$.

However, the dominant source of background is combinatorial, i.e., random combinations of $pK^-\pi^+$ that happen to have invariant mass within the $\Lambda_c^+$ signal peak.
A rough estimate of the level of this background can be obtained from published studies of 3-track $B$-meson decays where tag-$B$ reconstruction was performed. 
We use the study of $B^+\to \rho^0\ell^+ \nu$ conducted by Belle~\cite{Belle:2013hlo},
where Fig.~2 shows the background $\mmiss^2$ distribution for a data sample of $711~\textrm{fb}^{-1}$. 
The background is highest at $\mmiss^2\approx 2~\textrm{GeV}^2$, approximately $300$ events per $0.1~\textrm{GeV}^2$-wide bin. 
This corresponds to a $\sim0.35$~GeV-wide \mmiss\ bin around $m_\neu \approx  1.4$~GeV.
For simplicity, we also take $\sigma(\mmiss)\approx 0.35$~GeV at $m_\neu\approx 1.4$~GeV.
This estimate is based on Fig.~\ref{fig:res-BR-limit} above, but is conservative, since the resolution for three soft tracks is better than for the single hard proton used in Fig.~\ref{fig:res-BR-limit}.
Thus, the initial background estimate per typical $\sigma(\mmiss)$-wide bin is $300$ events for $711~\textrm{fb}^{-1}$, which is $21000$ per $50~\textrm{ab}^{-1}$. 
This is then modified by the following factors. 

First, the invariant mass of the $\pi^+\pi^-\ell^+$ system in Fig.~2 of Ref.~\cite{Belle:2013hlo} can take any value up to $m_B$, while that of the $pK^-\pi^+$ in our search must equal the known $\Lambda_c^+$ mass up to the $\Lambda_c^+$ resolution, which is about 5~MeV~\cite{BaBar:2006lxr}.
This leads to a background reduction of order $R_{\textrm{mass}} \approx 5~\textrm{MeV}/5~\textrm{GeV}=10^{-3}$.
Second, the area of the 
$\pi^+\pi^-\ell^+$ Dalitz plot in $B^+\to \rho^0\ell^+\nu$ can take values up to order $0.5\, m_B^2 \times 2 m_\rho \Gamma_\rho \approx 0.6~\textrm{GeV}^4$, while that of the decay $\Lambda_c^+\to pK^-\pi^+$ is of order $[(m_{\Lambda_c} - m_K)^2 - (m_p+m_\pi)^2] \times [(m_{\Lambda_c} - m_p)^2 - (m_K+m_\pi)^2] \approx 3~\textrm{GeV}^4$.
This increases the background estimate by 
$R_{\textrm{DP}} \approx 6$.
Lastly, we multiply by the ratio $R_{p/\ell}\approx 4\%/21\%$ of the proton and lepton abundances in $B$ decays and by the ratio $R_{K/\pi}\approx 13\%/180\%$ of the $K^-$ and $\pi^-$ abundances~\cite{Zyla:2020zbs}. 
Therefore, the final background estimate per $\sigma(\mmiss)$ region of missing mass is $1.7$ events at Belle~II.
We stress that this is to be taken only as an order of magnitude.
Nonetheless, it is similar to or somewhat larger than the background  estimated for \signal, which ranges between $0.25$ and $1.4$ events per $\sigma(\mmiss)$ bin.

In summary, relative to \signal, the mode \signalLambdaC\ has $R_\Br=6.3\%$ the branching fraction, $R_\epsilon=70\%$ the efficiency, a squared form factor in the range $R_{\textrm{FF}}=2-8\%$, and about the same  background level. 
Therefore, the limits on 
${\lambda''_{213}} \times (1 \ {\rm TeV}/{m_{\tilde q}})^2$ obtained from \signalLambdaC\ are expected to be about $\left(R_\Br R_\epsilon R_{\textrm{FF}} \right)^{-1/2} \approx 15-35$ times weaker than the ${\lambda''_{113}} \times (1 \ {\rm TeV}/{m_{\tilde q}})^2$ obtained from \signal.

Searching for \signalXiC\ involves reconstructing the $\Xi_c^+$ in a final state such as $\Xi^- \pi^+\pi^+$, which has a branching fraction of $2.9\pm 1.3\%$.
The $\Xi^-$ then decays to $\Lambda\pi^-$ 100\% of the time, followed by $\Lambda\to p\pi^-$ 64\% of the time. 
With five final-state tracks, the reconstruction efficiency is estimated to be about 
70\% that of \signalLambdaC.
The form factor for \signalXiC, shown in Fig.~\ref{fig:charmRatios}, is about $3/4$ that for \signalLambdaC.
Background considerations are similar to those of \signalLambdaC.
Since $\Br(B\to \Xi_c^+/\bar\Xi_c^- + \textrm{anything})\approx 1.6\%$~\cite{Zyla:2020zbs}, which is about 40\% of $\Br(B\to \Lambda_c^+/\bar \Lambda_c^-  + \textrm{anything})$, we estimate a corresponding reduction in background  from true $\Xi_c^+$.
The dominant background is again combinatorial, and is estimated to be about the same as for \signalLambdaC.
This comes from the fact that in inclusive $\Xi_c^+$ reconstruction the ratio of signal to combinatorial background is about $1/2$, as seen in Fig.~2 of Ref.~\cite{CLEO:1997kqj}, while a ratio of about $1$ is seen for inclusive $\Lambda_c^+$ reconstruction in Fig.~2 of Ref.~\cite{BaBar:2006lxr}.
In summary, we expect the 
limits on 
${\lambda''_{223}} \times (1 \ {\rm TeV}/{m_{\tilde q}})^2$ obtained from \signalXiC\ to be about $2.5$ times weaker than the ${\lambda''_{213}} \times (1 \ {\rm TeV}/{m_{\tilde q}})^2$ limits obtained from \signalLambdaC.

\section{Results for the RPV parameters}
\label{sec:results}

We proceed to calculate the sensitivity to the trilinear RPV couplings from the experimental sensitivity to the decay branching fractions of $B$ mesons into a charged baryon and a light neutralino.
As stated above, we take only one  RPV coupling to be non-vanishing at a time, either $\lambda''_{113}$ or $\lambda''_{123}$, and assume degenerate squark masses and no squark mixing.

\begin{figure}[h]
    \centering
    \includegraphics[width=0.6\textwidth]{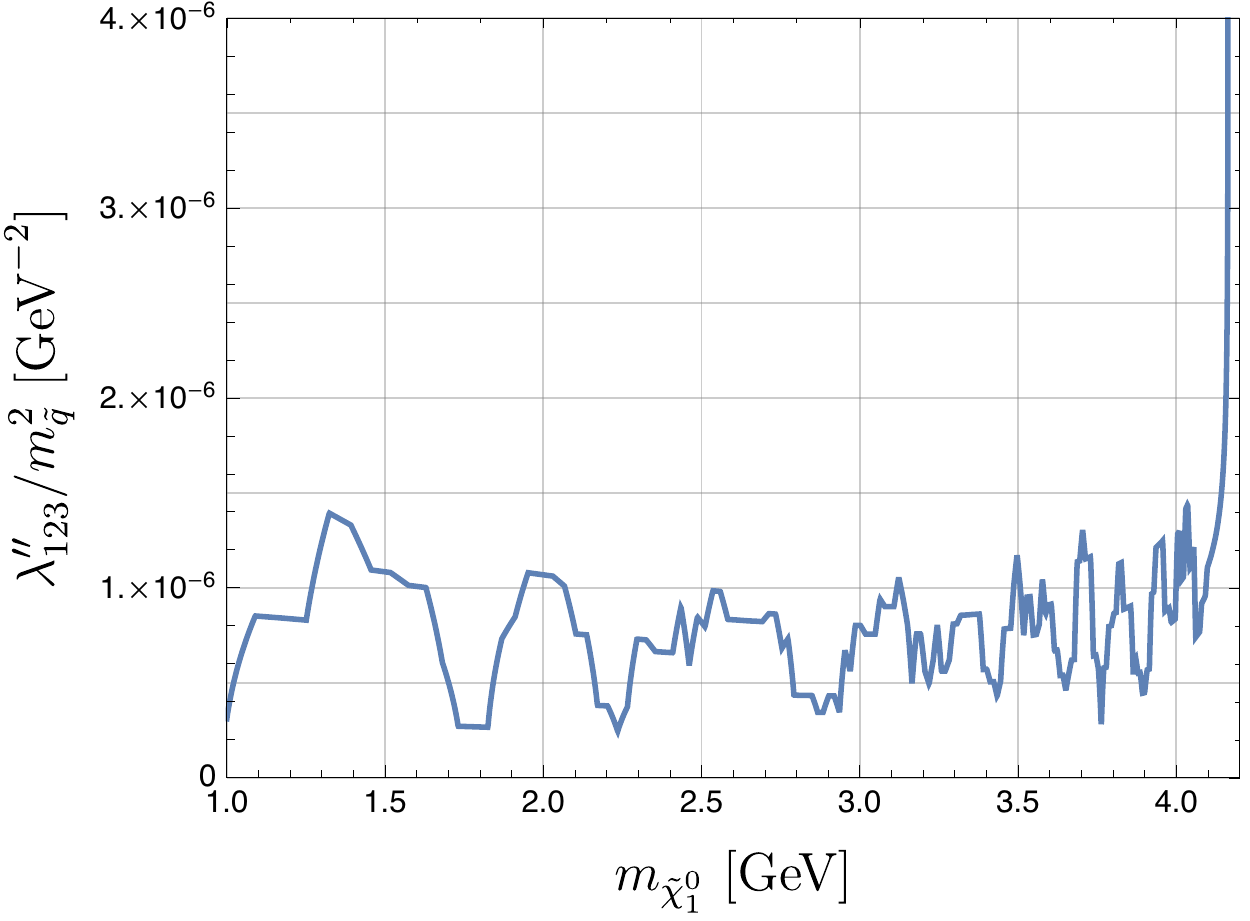}
    \caption{Limits on $\lambda''_{123}$ as a function of $m_{\neu}$ from the preliminary BABAR limits on $\Br(B^0\to \Lambda^0 \psi_D)$~\cite{BaBar:2023rer} reinterpreted in terms of $\signalLambda$.
    }
   \label{fig:babar-recast}
\end{figure}

In Fig.~\ref{fig:babar-recast} we show the limits on $\lambda''_{123}/m_{\tilde q}^2$ vs.~$m_{\neu}$ that we extract from the BABAR search for $B^0\to \Lambda^0 \psi_D$~\cite{BaBar:2023rer}.
These limits are around $\lambda''_{123}/m^2_{\tilde{q}} < 1\times 10^{-6}\,\mathrm{GeV}^{-2}$, with fluctuations that reflect the BABAR results. 
The BABAR search is effectively background-free in $m_{\neu}$ regions in which the \mmiss\ difference between background events is significantly larger than the experimental \mmiss\ resolution. 
With more than 100 times more data, the Belle~II search will probably not have a background-free region, unless further analysis improvements are implemented.
Therefore, the Belle~II sensitivity to $\lambda''_{123}$ is naively expected to be better than that of BABAR by about $1/\sqrt{R_L}$ for most regions of $m_{\neu}$.   
Related to this, as a result of the nonvanishing background, the Belle~II sensitivity is expected to become poorer as the \mmiss\ resolution decreases for low values of $m_{\neu}$.

\begin{figure}
    \centering
    \includegraphics[width=0.6\textwidth]{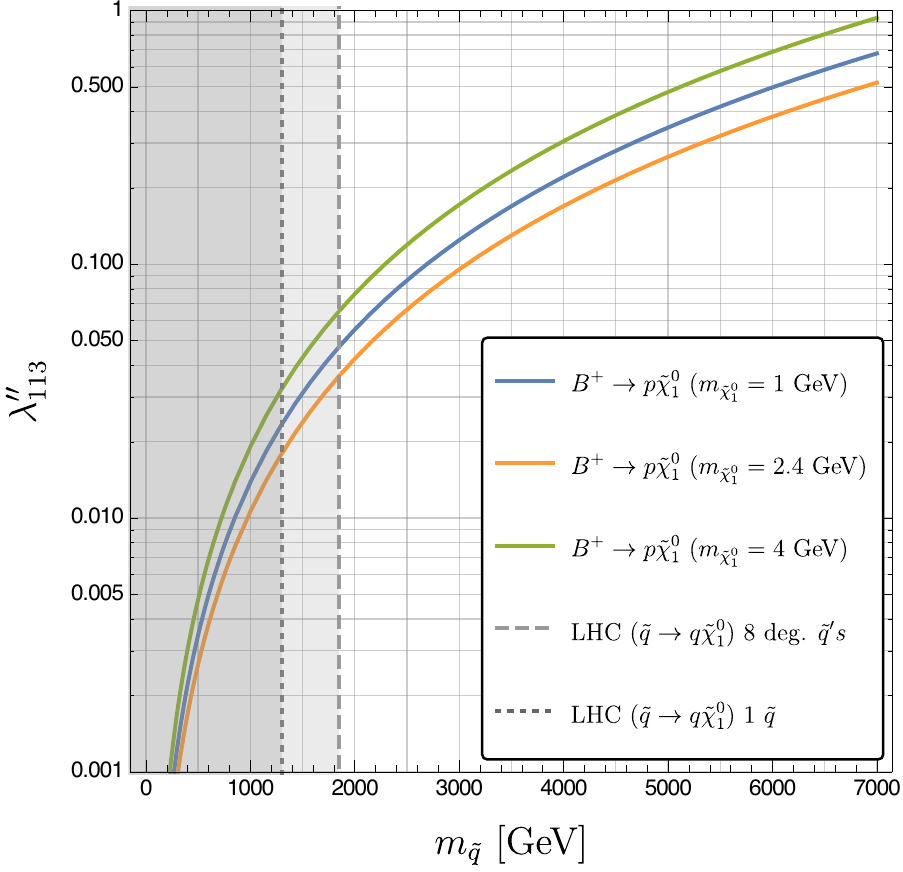}
\caption{
The expected upper bounds on
$\lambda''_{113}$ vs.~$m_{\tilde q}$ 
at Belle~II
for different values of the neutralino mass $m_{\neu}$, extracted from the proposed search for $B^+\to p \neu$  assuming the sensitivities given in
Sec.~\ref{sec:bgd}.
    We also plot current collider limits on the scalar squark masses
    under the hypotheses of eight mass-degenerate light-flavor squarks ($m_{\tilde q}\gtrsim 1850\mbox{ GeV}$) and only one light-flavor squark state  ($m_{\tilde q}\gtrsim 1300\mbox{ GeV}$) \cite{ATLAS:2020syg}.
    }
    \label{fig:limitlam113}
\end{figure}

Fig.~\ref{fig:limitlam113} shows the expected  sensitivity of the search for $B^+ \to p  \neu$ at Belle~II with a sample of $N_{BB} = 55\times 10^9$ $B\bar B$ pairs.  
The sensitivity is shown in the $\lambda''_{113}$ vs.~$m_{\tilde q}$ plane for the neutralino-mass values  $m_{\neu} = (1, 2.4, 4)$ GeV, and corresponds to $90\%$ confidence-level upper limits.
Fig.~\ref{fig:limitlam113} also compares our limits with current squark-mass limits, $m_{\tilde q} \gtrsim 1.85$ TeV for 8 mass-degenerate first- and second-generation squarks and $m_{\tilde q} \gtrsim 1.3$ TeV for a single squark belonging to the first or second generation.
These limits were obtained by the ATLAS collaboration from a search for $p p \to  \tilde q \tilde q$, $\tilde q \to q \neu$ in a signature of jets plus missing transverse momentum in a data sample with an integrated luminosity of $139~\text{fb}^{-1}$~\cite{ATLAS:2020syg}. 
The squark mass limits assume $\Br(\tilde q \to q \neu)=1$.
In our theoretical scenario this is a good approximation only for small values of $\lambda''_{113}$, for which the competing decay channel $\tilde q \to q q$ is suppressed.
As $\lambda''_{113}$ increases, the ATLAS $m_{\tilde q}$ limits weaken relative to the values shown in Fig.~\ref{fig:limitlam113}.
As mentioned in Sec.~\ref{sec:intro}, current bounds on $\lambda''_{113}$ and $\lambda''_{123}$ arising from di-nucleon and $n-\bar n$ oscillation are much weaker than bounds shown in Fig.~\ref{fig:limitlam113} and are hence not included in the figure. 

\begin{figure}
    \centering
    \includegraphics[width=0.6\textwidth]{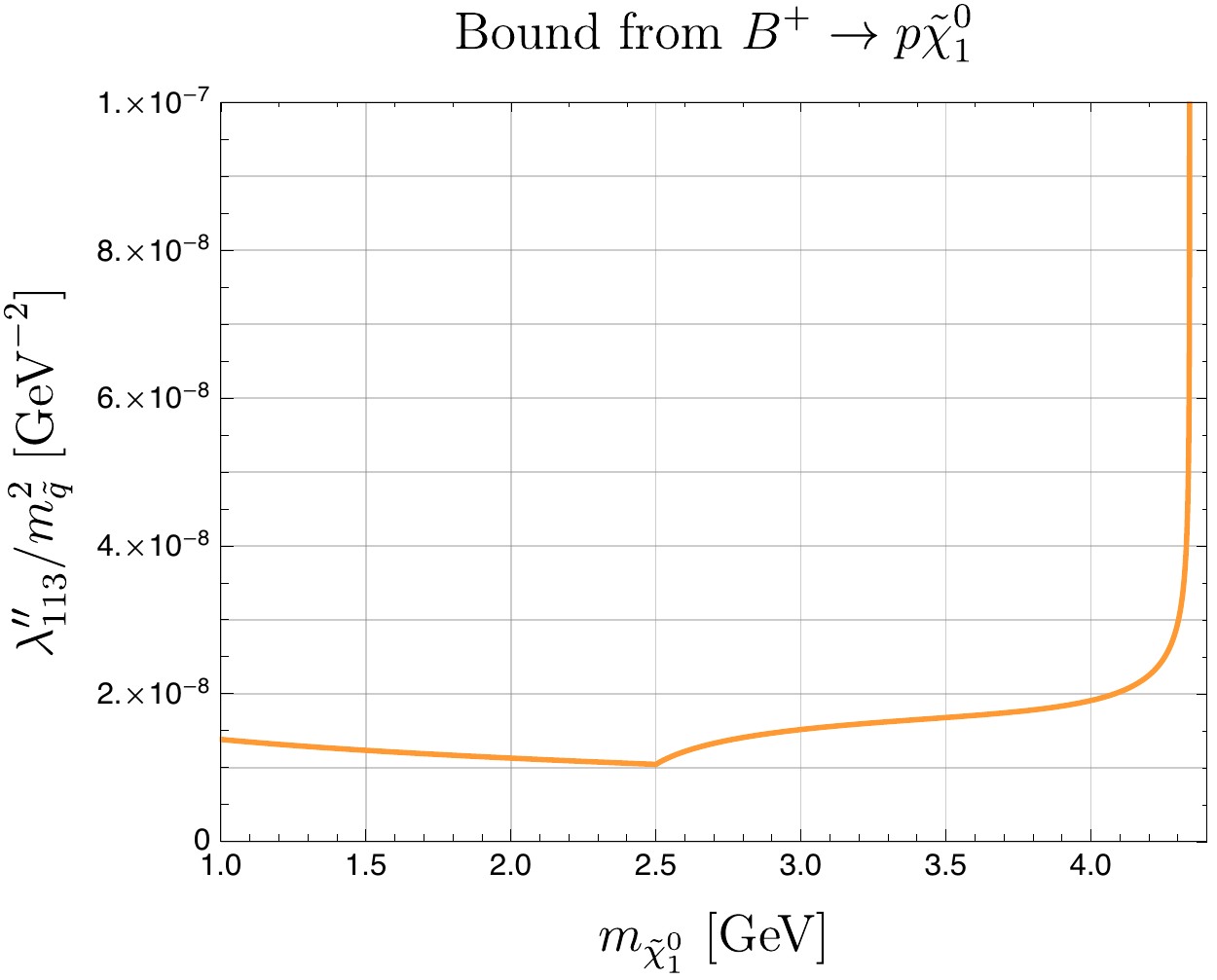}
    \caption{Upper bounds on $\lambda''_{113}/m_{\tilde q}^2$
    as a function of the neutralino mass, extracted from the process $B^+\to p \neu$,
    assuming the sensitivity given in Sec.~\ref{sec:bgd}.
    }
    \label{fig:limitlam113-2}
\end{figure}

Fig.~\ref{fig:limitlam113-2} shows our estimated Belle~II sensitivity in terms of bounds on $\lambda''_{113}/m_{\tilde q}^2$ vs. the neutralino mass.
The bound is between $1.5$ and $2\times 10^{-8}~\mathrm{GeV}^{-2}$ for most values of $m_{\neu}$.

\section{Discussion and Conclusions}
\label{sec:conclu}

A GeV-scale, Bino-like lightest neutralino $\neu$ is  allowed by current bounds, provided it decays,  e.g., via $R$-parity-violating (RPV) interactions, and is heavier than the proton, so as not to mediate rapid proton decay. 
In this work we consider such a light $\neu$ in the minimal realization of RPV supersymmetry. 
We propose to probe this scenario with a search for  $B$ meson decays into $\neu$ plus a baryon, specifically $p$, $\Lambda$, $\Sigma^+$, or $\Sigma^0$.
The decays are mediated by the RPV operators $\lambda''_{ijk}\bar{U}_i\bar{D}_j\bar{D}_k$. 
We study the case where only $\lambda''_{113}$ or $\lambda''_{123}$ is nonzero in the absence of squark mixing and with the simplifying assumption of mass-degenerate squarks.
We compute the corresponding $B$-meson decay rates in the framework of an effective phenomenological BNV Lagrangian.
We note that the lightest neutralino in our benchmark scenarios is so long-lived that it appears as missing energy at the detector level.

Experimentally, the presence of the $\neu$ and its mass can be determined from the missing mass at an $e^+ e^-$ $B$ factory, namely, BABAR, Belle, or Belle~II.
We interpret a BABAR search for $B^0$ decay into a $\Lambda$ baryon plus missing energy~\cite{BaBar:2023rer} to obtain current limits on $\lambda''_{123}/m_{\tilde q}^2$ around $1\times 10^{-6} / \textrm{GeV}^2$, where $m_{\tilde q}$ is the mediating squark mass. 
Relative to this search, we discuss the sensitivities of the modes $\signalSigma$ and $\signalSigmaZero$, as well as subtleties of the extrapolation to the final integrated luminosity of Belle~II.
Furthermore, we estimate the background level in a future analysis of $\signal$ and the resulting sensitivity to the branching fraction of this decay.
From this we obtain the sensitivity of Belle~II in terms of the upper limit on $\lambda''_{113}/m_{\tilde q}^2$, which is between $1.5$ and $2\times 10^{-8}~\mathrm{GeV}^{-2}$ for most kinematically allowed values of $m_{\neu}$.
For the currently allowed squark mass of about $1850$~GeV ($1300$~GeV) in the case of eight (one) degenerate squark states, this corresponds to bounds on $\lambda''_{113}$ as low as $0.05$ ($0.025$). 
The bounds are relevant also for multi-TeV squarks beyond the reach of direct detection at LHC.

We estimate the sensitivity to $\lambda''_{213}$ to be weaker than that of $\lambda''_{113}$ by a factor of $15-35$.
The sensitivity to $\lambda''_{223}$ is anticipated to be weaker still, by about $2.5$. 
Due to lack of sufficient experimental information, these estimates are to be taken as order of magnitude only. 
Nonetheless, they demonstrate that the proposed searches for \signalLambdaC\ and \signalXiC\ can provide higher sensitivities than the bounds obtained from dinuclen decays for the scenario under study here.

Lastly, we note that similar studies can be performed with other processes, exploiting the capabilities of other experiments.
At LHCb, one can use the decay $B^+\to p \pi^+\pi^-\neu$, with the $B^+$ produced in the decay $B^{*0}_{s2}\to B^+ K^-$, which has already been used to study semileptonic $B$ decays~\cite{LHCb:2018azb}. 
Owing to the unobserved neutralino, the 4-momenta of the $\neu$, $B^+$, and $B^{*0}_{s2}$ are unknown.
However, they can be determined up to a 2-fold ambiguity from the 12 constraints of the signal decay chain: 4-momentum conservation in the $B^{*0}_{s2}$ and $B^+$ decays, the known $B^{*0}_{s2}$ and $B^+$ masses, and the flight direction of the $B^+$ determined from the position of the $p \pi^+\pi^-$ vertex.
Solution of the constraint equations results in two values of the neutralino mass, $m_\pm$, similarly to Refs.~\cite{Dib:2019tuj, Dey:2020juy}. 
Signal events peak more significantly than background in the $m_\pm$ distribution, except for expected peaks around masses of (mostly charmed) baryons that decay to only neutral particles.
The feasibility of this approach depends, among other factors, on the $m_\pm$ measurement resolution.
The decay chains
$B^{*+}_{s2}\to B^0 K^+$, $B^0\to p \pi^-\neu$ and $\Sigma_b^+\to \Lambda_b \pi^+$, $\Lambda_b\to \pi^+\pi^- \neu$ may be used in a similar way.
We note that at $e^+e^-$ $B$ factories, $B^{*+}_{s2}$ production is not feasible, and final staes with additional pions are probably somewhat disadvantageous, having lower efficiency and potentially higher combinatorial background than single-baryon final states.
One may also probe $\lambda''_{212}$ in the decays $D^+\to \overline{\Sigma^-} \neu$ and $D^0\to \bar\Lambda^0 \neu$ for a smaller range of the neutralino mass. 
Specifically, the $D$ can be produced in $\psi(3770)\to D \bar D$ at BESIII, with full reconstruction of the $\bar D$ in a manner identical to the Belle~II analysis we propose here.

\section*{Acknowledgements:} 
We thank Florian Domingo and Martin Hirsch for useful discussions, and Dexu Lin and the BABAR collaboration for help in the interpretation of Ref.~\cite{BaBar:2023rer}.
Z. S. W. is supported by the Ministry of Science and Technology (MoST) of Taiwan with grant number MoST-110-2811-M-007-542-MY3.
C.D. acknowledges support from FONDECYT (Chile) Grant No. 1210131, and ANID (Chile) PIA/APOYO AFB180002 and AFB220004.
J. C. H.  acknowledges support from grant FONDECYT (Chile) No.1201673  and ANID – Programa Milenio - code ICN2019\_044. 
V. E. L. is supported by by BMBF (Germany) ``Verbundprojekt 05P2021 (ErUM-FSP T01) -
Run 3 von ALICE am LHC: Perturbative Berechnungen von Wirkungsquerschnitten
f\"ur ALICE'' (F\"orderkennzeichen: 05P21VTCAA), by ANID (Chile) PIA/APOYO AFB180002 and AFB220004, by FONDECYT (Chile) under Grant No. 1191103, 
and by ANID$-$Millen\-nium Program$-$ICN2019\_044 (Chile).
N. A. N. is supported by ANID (Chile) under the grant ANID REC Convocatoria Nacional Subvenci\'on a Instalaci\'on en la Academia Convocatoria A\~no 2020, PAI77200092.
A. S. is supported by grants from the Israel Science Foundation, the United States-Israel
Binational Science Fund, and the Tel
Aviv University Center for Artificial Intelligence and Data Science.

\appendix 
\section{Transition Form Factors} 
\label{app:formfactors} 

Here, we list the results for the form factors of the $B^+ \to p (\Sigma^+)  \neu$ and 
$B^+ \to \Lambda^0 (\Sigma^0)  \neu$
transitions and  discuss how to extract the results for the couplings of light baryons with corresponding 
three-quark currents. 
First, we specify the nucleon couplings $\alpha$ and $\beta$, which define the matrix elements of three-quark operators  
between nucleon and vacuum (they are essential ingredient of our calculation of the transition form factors): 
\eq 
\la p|{\cal O}_{uud}^{LR}|0\ra =   \bar u_p(p',s') \, \alpha \, P_R \,, \quad 
\la p|{\cal O}_{uud}^{LL}|0\ra = - \bar u_p(p',s') \, \beta  \, P_L \,, \label{eq:3quarkops}
\en 
where 
\eq
{\mathcal O}_{q_1q_2q_3}^{XY} &=& \varepsilon^{a_1a_2a_3} \,  
\left(\bar q_3^{a_3} P_X \bar q_2^{a_2}\right) \bar q_1^{a_1} P_Y \,, 
\nonumber\\
\overline{{\mathcal O}}_{q_1q_2q_3}^{XY} &=& \varepsilon^{a_1a_2a_3} \,  
P_Y q_1^{a_1} \left(q_2^{a_3} P_X q_3^{a_3}\right),    
\en
and $X,Y = L,R$. In our calculations we deal only with ${\mathcal O}_{q_1q_2q_3}^{LL}$ operators.
Therefore, we need only $\beta$ couplings and form factors 
$W_{0,1}^{LL}(q^2)$. In addition, for completeness, we also discuss and show results for the 
case of the ${\mathcal O}_{q_1q_2q_3}^{LR}$ operators, 
i.e. for $\alpha$ couplings and $W_{0,1}^{LR}(q^2)$ form factors. 

The nucleon couplings $\alpha$ and $\beta$ 
have been calculated using various QCD-motivated approaches
and lattice QCD (see, e.g., detailed compilation of the predictions 
in Table 5 of Ref.~\cite{Aoki:2008ku}
and recent lattice results in Ref.~\cite{Yoo:2021gql}). One can see that results of QCD models
and lattice QCD for the $\alpha$ coupling range from 
$-0.003$~GeV$^3$ to $-0.03$~GeV$^3$ and 
from $-0.006$~GeV$^3$ to $-0.03$~GeV$^3$, 
respectively. In our analysis,
we take recent lattice result $\alpha = - 0.01257 \simeq - 0.0126$~GeV$^3$ of the USQCD Collaboration~\cite{Yoo:2021gql} 
(continuum extrapolation) as central value and take into account its variation as: 
$\alpha = -0.0126^{+0.0174}_{-0.0096}$~GeV$^3$. 
In case of the $\beta$ coupling mainly lattice QCD results are available, which are distributed from 
0.01~GeV$^3$ to 0.0127~GeV$^3$. 
Therefore, for the $\beta$ coupling we will use the following predictions with lattice QCD 
approaches, of the central value from a recent paper of the USQCD Collaboration~\cite{Yoo:2021gql} 
and take into account error occurring in other lattices calculations~\cite{Aoki:2008ku,Yoo:2021gql}:  
$\beta = 0.0108 \pm 0.007$~GeV$^3$. 

For the case of the transition $B^+ \to \Sigma^+  \neu$ 
we need the $\Sigma^+$ coupling $\beta_{\Sigma^+}$ defining the matrix element 
$\la \Sigma^+|{\cal O}_{uus}^{LL}|0\ra$ (for completeness we 
also calculate the coupling $\alpha_{\Sigma^+}$ defining the matrix element 
$\la \Sigma^+|{\cal O}_{uus}^{LR}|0\ra$): 
\eq\label{alphabetaSigma}  
\la \Sigma^+|{\cal O}_{uus}^{LR}|0\ra =   \bar u_{\Sigma^+}(p',s') \, \alpha_{\Sigma^+} \, P_R \,, \quad 
\la \Sigma^+|{\cal O}_{uus}^{LL}|0\ra = - \bar u_{\Sigma^+}(p',s') \, \beta_{\Sigma^+}  \, P_L \,.
\en
The couplings $\alpha_{\Sigma^+}$ and $\beta_{\Sigma^+}$ can be fixed using results for the so-called 
vector $\lambda_V$ and tensor $\lambda_T$ couplings estimated  
in Ref.~\cite{Liu:2008yg} using QCD sum rules: 
\eq\label{couplingsVT}  
\la 0|(suu)_V|\Sigma^+\ra = \lambda^V_{\Sigma^+} \, u_{\Sigma^+}(p,s) \,,  \quad 
\la 0|(suu)_T|\Sigma^+\ra = \lambda^T_{\Sigma^+} \, u_{\Sigma^+}(p,s) \,, 
\en
where 
\eq\label{currentsVT} 
(q_1q_2q_3)_V &=& \varepsilon^{a_1a_2a_3} \, \gamma^\mu \gamma^5 q_1^{a_1} \, 
(q_2^{a_2} C \gamma_\mu q_3^{a_3})\,, \nonumber\\
(q_1q_2q_3)_T &=& \varepsilon^{a_1a_2a_3} \, \sigma^{\mu\nu} \gamma^5 q_1^{a_1} \, 
(q_2^{a_2} C \sigma^{\mu\nu} q_3^{a_3}) 
\en 
are the vector $V$ and tensor $T$ three-quark currents (operators).  
Also we will need for further calculations the pseudoscalar $P$ and scalar $S$ three-quark 
operators: 
\eq\label{currentsPS} 
(q_1q_2q_3)_P &=& \varepsilon^{a_1a_2a_3} \, q_1^{a_1} \, 
(q_2^{a_2} C \gamma^5 q_3^{a_3})\,, \nonumber\\
(q_1q_2q_3)_S &=& \varepsilon^{a_1a_2a_3} \, \gamma^5 q_1^{a_1} \, 
(q_2^{a_2} C q_3^{a_3}).
\en 
Using Fierz trasformation relating $V$ and $T$ currents defined in Eq.~(\ref{currentsVT}) 
we can relate them with pseudoscalar $P$ and scalar $S$ currents and then with the currents 
with specific chirality occurring in Eq.~(\ref{alphabetaSigma}). In particular, 
for three-quark operators with flavor content of the $\Sigma^+$ baryon we have the 
following Fierz identities: 
\eq\label{Fierz_suu} 
(suu)_V = 2 \Big[ (uus)_P - (uus)_S \Big] \,, \quad 
(suu)_T = 4 \Big[ (uus)_P + (uus)_S \Big] \,.
\en 
Next using  
\eq\label{Fierz_uus} 
(uus)_P &=& \overline{{\cal O}}_{uus}^{RR} 
          - \overline{{\cal O}}_{uus}^{LL} 
          + \overline{{\cal O}}_{uus}^{RL} 
          - \overline{{\cal O}}_{uus}^{LR}
\,, \\
(uus)_S &=& \overline{{\cal O}}_{uus}^{RR} 
          - \overline{{\cal O}}_{uus}^{LL} 
          - \overline{{\cal O}}_{uus}^{RL} 
          + \overline{{\cal O}}_{uus}^{LR}
\,,\en 
we can express the matrix elements of $V$ and $T$ currents through the 
matrix elements of the $\overline{{\cal O}}_{uus}^{RR}$ operators and, therefore, 
relate then the pair of the couplings $\lambda^V_{\Sigma^+},\lambda^T_{\Sigma^+}$ 
with couplings $\alpha_{\Sigma^+}$ and $\beta_{\Sigma^+}$:
\eq\label{VT_Operators_SP}
\la 0|(suu)_V|\Sigma^+\ra &=& 
4  \biggl[ \la 0|\overline{{\cal O}}_{uus}^{RL} |\Sigma^+\ra 
         - \la 0|\overline{{\cal O}}_{uus}^{LR} |\Sigma^+\ra 
\biggr] \nonumber\\
&=& 4 \alpha_{\Sigma^+} \, u_{\Sigma^+}(p,s) \,,  \\
\la 0|(suu)_T|\Sigma^+\ra &=& 
8  \biggl[ \la 0|\overline{{\cal O}}_{uus}^{RR} |\Sigma^+\ra 
         - \la 0|\overline{{\cal O}}_{uus}^{LL} |\Sigma^+\ra 
\biggr]  \nonumber\\
&=& - 8 \beta_{\Sigma^+} \, u_{\Sigma^+}(p,s) \,. 
\label{VT_Operators_SP2}
\en
Matching Eqs.~(\ref{couplingsVT}), 
(\ref{VT_Operators_SP}), and~(\ref{VT_Operators_SP2})
we arrive at the required relations: 
\eq
\alpha_{\Sigma^+} =   \frac{1}{4} \lambda^V_{\Sigma^+} \,, \quad \
\beta_{\Sigma^+}  = - \frac{1}{8} \lambda^T_{\Sigma^+} \,.
\en 
Taking $\lambda^{V,T}_{\Sigma^+}$ couplings from Ref.~\cite{Liu:2008yg} 
with the sign conventions as in Ref.~\cite{Aoki:2008ku} we get: 
\eq
\alpha_{\Sigma^+} = - (0.743 \pm 0.030) \times 10^{-2} \ {\rm GeV}^3 \,, 
\quad 
\beta_{\Sigma^+} =  (0.654 \pm 0.030) \times 10^{-2} \ {\rm GeV}^3 \,.
\en 

For the transitions $B^0 \to \Lambda^0  \neu$ and  $B^0 \to \Sigma^0  \neu$ 
we need the $\Lambda^0$ and $\Sigma^0$ couplings $\beta_{\Lambda^0}$ and $\beta_{\Sigma^0}$ 
defining the following matrix elements 
\eq 
\la \Lambda^0|{\cal O}_{sud}^{LL}|0\ra = - \bar u_p(p',s') \, \beta_{\Lambda^0}  \, P_L \,, 
\\
\la \Sigma^0|1/\sqrt{2} [{\cal O}_{uds}^{LL} + 
{\cal O}_{dus}^{LL}] |0\ra 
= - \bar u_p(p',s') \, \beta_{\Sigma^0}  \, P_L \,.
\en 
The $\beta_{\Lambda^0}$ coupling was calculated in Ref.~\cite{Liu:2008yg} using QCD sum rules: 
\eq
\beta_{\Lambda^0} =  (0.926 \pm 0.056) \times 10^{-2} \ {\rm GeV}^3 \,.
\en 
As for $\beta_{\Sigma^0}$ we accept isospin-symmetry relation: 
$\beta_{\Sigma^0} = \beta_{\Sigma^+}$.

As we stressed above, the form factors $W_0^{LX}(q^2)$ and $W_1^{LX}(q^2)$ 
defining matrix elements $\la p| {\mathcal O}^{LX} |B^+\ra$ can be calculated using  
SU(5) extension of the SU(3) version of the phenomenological Lagrangian proposed 
and used in Refs.~\cite{Gavela:1988cp} 
and~\cite{JLQCD:1999dld,Aoki:2006ib,Aoki:2008ku,Aoki:2013yxa,Yoo:2021gql}. 
At the leading order 
form factors $W_0^{LX}(q^2)$ and $W_1^{LX}(q^2)$ are contributed by 
direct BNV couplings of proton and pseudoscalar mesons and 
pole contributions because of exchange of baryon resonances. 
In particular, using results 
for the BNV matrix elements $\la p| {\mathcal O}^{LX} |K^+\ra$ and making 
replacements $f_K \to f_B$, $m_{\Lambda^0} \to m_{\Lambda_b}$, and 
$m_{\Sigma} \to m_{\Sigma_b}$, one obtains: 
\eq\label{W0LL}  
W_0^{LL}(q^2) &=& - \frac{\beta}{f_B} \, 
\biggl[ g^{\tilde bR} \, \left(1 + \dfrac{D+3F}{3} \, 
\dfrac{m_{\Lambda_b} m_p + q^2}{m_{\Lambda_b}^2 - q^2}\right) 
\nonumber\\
&+& g^{\tilde uR} \, \left(- 1 
- \dfrac{D+3F}{6} \, \dfrac{m_{\Lambda_b} m_p + q^2}{m_{\Lambda_b}^2 - q^2}   
+ \dfrac{D-F}{2}  \, \dfrac{m_{\Sigma_b} m_p + q^2}{m_{\Sigma_b}^2 - q^2}\right) 
\nonumber\\
&+& g^{\tilde dR}  \, \left( 
  \dfrac{D+3F}{6} \, \dfrac{m_{\Lambda_b} m_p + q^2}{m_{\Lambda_b}^2 - q^2}  
+ \dfrac{D-F}{2}  \, \dfrac{m_{\Sigma_b} m_p + q^2}{m_{\Sigma_b}^2 - q^2}\right) 
\biggr] \,, 
\en 
\eq\label{W1LL}  
W_1^{LL}(q^2) &=& - \frac{\beta}{f_B} \, m_{\neu} \,  
\biggl[ g^{\tilde bR} \, \dfrac{D+3F}{3} \, 
\dfrac{m_{\Lambda_b} + m_p}{m_{\Lambda_b}^2 - q^2}  
\nonumber\\
&+& g^{\tilde uR} \, \left(
- \dfrac{D+3F}{6} \, \dfrac{m_{\Lambda_b} + m_p}{m_{\Lambda_b}^2 - q^2} 
+ \dfrac{D-F}{2}  \, \dfrac{m_{\Sigma_b}  + m_p}{m_{\Sigma_b}^2 - q^2}\right) 
\nonumber\\
&+& g^{\tilde dR}  \, \left( 
  \dfrac{D+3F}{6} \, \dfrac{m_{\Lambda_b} + m_p}{m_{\Lambda_b}^2 - q^2} 
+ \dfrac{D-F}{2}  \, \dfrac{m_{\Sigma_b} + m_p}{m_{\Sigma_b}^2 - q^2}\right) 
\biggr] \,, 
\en 
\eq\label{W0LR} 
W_0^{LR}(q^2) &=& - \frac{\alpha}{f_B} \, 
\biggl[ g^{\tilde bL} \, \left(1 + \dfrac{D+3F}{3} \, 
\dfrac{m_{\Lambda_b} m_p + q^2}{m_{\Lambda_b}^2 - q^2}\right) 
\nonumber\\
&+& g^{\tilde uL} \, \left(- 1 
- \dfrac{D+3F}{6} \, \dfrac{m_{\Lambda_b} m_p + q^2}{m_{\Lambda_b}^2 - q^2}   
+ \dfrac{D-F}{2}  \, \dfrac{m_{\Sigma_b} m_p + q^2}{m_{\Sigma_b}^2 - q^2}\right) 
\nonumber\\
&+& g^{\tilde dL}  \, \left( 
  \dfrac{D+3F}{6} \, \dfrac{m_{\Lambda_b} m_p + q^2}{m_{\Lambda_b}^2 - q^2}   
+ \dfrac{D-F}{2}  \, \dfrac{m_{\Sigma_b} m_p + q^2}{m_{\Sigma_b}^2 - q^2}\right) 
\biggr] \,, 
\en 
\eq\label{W1LR}  
W_1^{LR}(q^2) &=& - \frac{\alpha}{f_B} \, m_{\neu} \,  
\biggl[ g^{\tilde bL} \, \dfrac{D+3F}{3} \, 
\dfrac{m_{\Lambda_b} + m_p}{m_{\Lambda_b}^2 - q^2} 
\nonumber\\
&+& g^{\tilde uL} \, \left(
- \dfrac{D+3F}{6} \, \dfrac{m_{\Lambda_b} + m_p}{m_{\Lambda_b}^2 - q^2}  
+ \dfrac{D-F}{2}  \, \dfrac{m_{\Sigma_b}  + m_p}{m_{\Sigma_b}^2 - q^2}\right) 
\nonumber\\
&+& g^{\tilde dL}  \, \left( 
  \dfrac{D+3F}{6} \, \dfrac{m_{\Lambda_b} + m_p}{m_{\Lambda_b}^2 - q^2}  
+ \dfrac{D-F}{2}  \, \dfrac{m_{\Sigma_b} + m_p}{m_{\Sigma_b}^2 - q^2}\right) 
\biggr] \,, 
\en 
where $D=0.8$, $F=0.47$, 
$f_B = 0.192$ GeV, 
$m_{\Lambda_b} = 5.6196$   GeV, 
$m_{\Sigma_b}  = 5.81056$  GeV, 
$m_{B^+}       = 5.27934$  GeV, and
$m_p           = 0.93827$ GeV.
In case of the $B^+ \to \Sigma^+ + \neu$ transition we 
replace the masses of the baryons in the expressions for the form 
factors (see Eqs.~(\ref{W0LL}) and~(\ref{W1LL})) and kinematical formula for the 
decay rate~(Eq.~\eqref{decay_width}) accordingly: 
$m_{\Lambda_b} \to m_{\Xi_b} = 5.7919$ GeV, 
$m_{\Sigma_b} \to m_{\Xi_b'} = 5.93502$ GeV, and  
$m_p \to m_{\Sigma^+} = 1.18937$ GeV. 

Note that direct BNV couplings of proton/$\Sigma^+$ and $B^+$ meson occur on the 
partonic level only in the case of the subprosses induced by the $\tilde b$ and $\tilde u$ squarks and 
can not occur in the case of the $\tilde d$ squark for the proton mode and 
$\tilde s$ for the $\Sigma^+$ mode. 
In particular, direct BNV coupling of proton and $\Sigma^+$ with $B^+$ means that 
the latter transits into two three-quark currents. One current corresponds to the proton current $J^P$ 
or $\Sigma^+$ hyperon current $J^{\Sigma^+}$ and 
another one to the BNV operator $J^{\rm BNV}$. The BNV operator has 
a form of scalar diquark $[q_1q_2]_0$ coupled to the third quark $q_3$. 
We have three possibilities for the $J^{\rm BNV}$ 
(for simplicity we drop color indices): 
\eq 
J^{\rm BNV}_{bud} = \bar b [\bar u \bar d]_0\,, \quad 
J^{\rm BNV}_{udb} = \bar u [\bar d \bar b]_0\,, \quad 
J^{\rm BNV}_{dub} = \bar d [\bar u \bar b]_0 \,,
\en 
for the proton mode and 
\eq 
J^{\rm BNV}_{bus} = \bar b [\bar u \bar s]_0\,, \quad 
J^{\rm BNV}_{usb} = \bar u [\bar s \bar b]_0\,, \quad 
J^{\rm BNV}_{sub} = \bar s [\bar u \bar b]_0 \,,
\en 
for the $\Sigma^+$ mode. 
It means that the $B^+$ meson having spin $J=0$ can decay into light baryons having 
scalar quark configuration.
In our case, proton and $\Sigma^+$ hyperon could be formed as quark-scalar diquark bound state: 
\eq 
J^p = u [ud]_0\, \quad J^{\Sigma^+} = u [us]_0 \,. 
\en 
Therefore, the following direct transitions are allowed: (1) $B^+$ meson 
annihilates into $u$ and $\bar b$ quarks, which further play the role of third quarks in 
light baryon and BNV current. In this case, the pair of scalar diquark $[u d]_0$ and  
antidiquark $[\bar u \bar d]_0$ is produced from vacuum; 
(2) $B^+$ meson annihilates into the pair of scalar diquark $[u d]_0$ and 
antidiquark $[\bar b \bar d]_0$. The allowed direct modes are listed as: 

Direct transition induced by the $\tilde b$ squark 
\eq 
B^+(u \bar b) \to J^p + J^{\rm BNV}_{bud} \,, \quad 
B^+(u \bar b) \to J^{\Sigma^+} + J^{\rm BNV}_{bus} \,.
\en 

Direct transition induced by the $\tilde u$ squark  
\eq 
B^+(u \bar b) \to J^p + J^{\rm BNV}_{udb} \,, \quad 
B^+(u \bar b) \to J^{\Sigma^+} + J^{\rm BNV}_{usb} \,.
\en 

We have no direct transitions induced by the $\tilde d$ squark for 
the proton mode and by the $\tilde s$ squark for the $\Sigma^+$ mode, 
because it requires production of the pair of scalar diquark 
contributing to the $J^{\rm BNV}$ current and 
vector diquark ($d \{uu\}_1$ or $s \{uu\}_1$) contributing 
to the current of light baryon, which is suppressed 
by spin conservation. 

Now we turn to discussion of the allowed pole transitions induced by bottom baryon 
resonances. We have two types of bottom baryon resonances with quark content of 
the BNV operators: baryons $\Lambda_b$ and $\Xi_b$ with antisymmetric light quark 
configuration, whose three-quark currents could 
have a form of light spin-0 diquark $[ud]_0$ and $[us]_0$, respectively, coupled to 
the $b$ quark 
\eq\label{Lambda_Xi_currents1} 
J^{\Lambda_b} = b [ud]_0  \,,  \quad 
    J^{\Xi_b} = b [us]_0 \,,
\en 
or vector diquarks composed of 
light quark $u$, $d$, or $s$ and bottom quark $b$: 
\eq 
J^{\Lambda_b} = u \{db\}_1 - d \{ub\}_1 \,,  \quad 
    J^{\Xi_b} = u \{sb\}_1 - s \{ub\}_1 \,,
\en 
and baryons $\Sigma_b=b \{ud\}_1$ and $\Xi'_b=b \{us\}_1$ with symmetric light quark 
configuration, whose three-quark currents  
could have a form of light spin-1 diquark $\{ud\}_1$ and $\{us\}_1$, respectively 
coupled to the $b$ quark 
\eq\label{Sigma_Xi_currents1} 
J^{\Sigma_b} = b \{ud\}_1  \,,  \quad 
  J^{\Xi'_b} = b \{us\}_1 \,,
\en 
or scalar diquarks composed of light quark $u$, $d$, or $s$ 
and bottom quark $b$:
\eq\label{Sigma_Xi_currents2} 
J^{\Sigma_b} = u [db]_0 + d [ub]_0 \,,  \quad 
  J^{\Xi'_b} = u [sb]_0 + s [ub]_0  \,. 
\en 
Therefore, pole transitions with $\Lambda_b$ and $\Xi_b$ baryon resonances 
are allowed for all three squarks $\tilde b$, $\tilde u$, and $\tilde d$ 
in the case of the proton mode and  $\tilde b$, $\tilde u$, and $\tilde s$ 
in the case of the $\Sigma^+$ mode. 

On the partonic level 
$B^+ \to p  \Lambda_b$ and $B^+ \to \Sigma^+  \Xi_b$
transitions are induced by the quark transitions, respectively:   
\eq 
B^+(u \bar b) \to p(u [ud]_0) 
+ \bar\Lambda_b^0(\bar b [\bar u \bar d]_0) \,, \quad 
B^+(u \bar b) \to \Sigma^+(u [us]_0) 
+ \bar\Xi_b^0(\bar b [\bar u \bar s]_0) \,. 
\en 
Next, the $\Lambda_b$ or $\Sigma_b$ converses into 
$J^{\rm BNV}$ with emission of the neutralino. In particular, 
the transitions induced by the squark
$\tilde b$ exchange are
\eq\label{b_channel_LbXb1}  
\bar b [\bar u \bar d]_0 \to \bar b [\bar u \bar d]_0 
\,, \quad 
\bar b [\bar u \bar s]_0 \to \bar b [\bar u \bar s]_0 
\,. 
\en 
Transitions induced by the squarks
$\tilde u$ and $\tilde d(s)$ are
\eq\label{u_channel_LbXb1}  
\bar u [\bar d \bar b]_0 \to \bar u [\bar d \bar b]_0 \,, \quad 
\bar u [\bar s \bar b]_0 \to \bar u [\bar s \bar b]_0 \,,
\en 
and 
\eq\label{ds_channel_LbXb1} 
\bar d [\bar u \bar b]_0 \to \bar d [\bar u \bar b]_0 \,, \quad 
\bar s [\bar u \bar b]_0 \to \bar s [\bar u \bar b]_0 \,.
\en 
One can explain relative contributions of 
all three channels to the hadronic form factors. 
The $\tilde u$ and $\tilde d(s)$
channel contributions are generated by the Fierz 
transformation of the $\Lambda_b$ and $\Xi_b$ currents 
$J^{\Lambda_b} = b [ud]_0$ and $J^{\Xi_b} = b [us]_0$
allowing to interchange $b \leftrightarrow u$ and 
$b \leftrightarrow d(s)$. It leads to the currents 
\eq 
J^{\Lambda_b} &=& - \frac{1}{2} \, u [db]_0 + \ldots = \frac{1}{2} \, d [ub]_0 + \ldots 
\,, 
\label{uds_channel_LbXb1} \\
J^{\Xi_b} &=& - \frac{1}{2} \, u [sb]_0 + \ldots = \frac{1}{2} \, s [ub]_0 + \ldots \,.
\label{uds_channel_LbXb2} 
\en 
From Eqs.~(\ref{b_channel_LbXb1})-(\ref{uds_channel_LbXb2}) 
one can see that relative contributions of the 
$\tilde b$, $\tilde u$, and $\tilde d(s)$ squark channels  
are $2 : (-1) : 1$. Note, that the contributions 
of $\tilde u$ and $\tilde d(s)$ squarks are equal 
on magnitude and of opposite sign as a result of antisymmetric configuration of $(u,d)$ and $(u,s)$ quarks in the $\Lambda_b$ and $\Xi_b$ baryons, respectively (see Eq.~(\ref{Lambda_Xi_currents1})). 

From Eqs.~(\ref{Sigma_Xi_currents1}) and~(\ref{Sigma_Xi_currents2}), we conclude that in the case of the pole contributions induced by the bottom baryons $\Sigma_b$ and $\Xi'_b$ with symmetric configurations of light quarks, the squark $\tilde b$ mode is absent because spin-1 diquark $[ud(s)]_1$ cannot transit to the spin-0 diquark $[ud(s)]_0$. 
On the other hand, the $\tilde u$ and $\tilde d(s)$ squark modes are allowed and they have exactly the same contributions because of the symmetric configuration of light quarks in the $\Sigma_b$ and $\Xi'_b$ baryons. 

By analogy, we consider, in addition, decay modes of the neutral $B^0$ mesons 
$B^0\to \Lambda^0 \neu$ and $B^0\to \Sigma^0 \neu$. In particular, 
the corresponding hadronic form factors defining these decay read: 
\eq\label{W0LL1}  
W_0^{LL}(q^2) &=& -  \sqrt{\dfrac{3}{2}} \, \frac{\beta_{\Lambda^0}}{f_B} \, 
\biggl[ g^{\tilde bR} \, \left(-\dfrac{1}{2} - \dfrac{D+3F}{6} \, 
\dfrac{m_{\Xi_b} m_{\Lambda^0} + q^2}{m_{\Xi_b}^2 - q^2}\right) 
\nonumber\\
&+& g^{\tilde uR} \, \left(- \dfrac{1}{2} 
+ \dfrac{D+3F}{12} \, \dfrac{m_{\Xi_b} m_{\Lambda^0} + q^2}{m_{\Xi_b}^2 - q^2}   
+ \dfrac{3 (D-F)}{4}  \, \dfrac{m_{\Xi'_b} m_{\Lambda^0} + q^2}{m_{\Xi'_b}^2 - q^2}\right) 
\nonumber\\
&+& g^{\tilde sR}  \, \left( - 1
  - \dfrac{D+3F}{12} \, \dfrac{m_{\Xi_b} m_{\Lambda^0} + q^2}{m_{\Xi_b}^2 - q^2}  
+ \dfrac{3 (D-F)}{4}   \, \dfrac{m_{\Xi'_b} m_{\Lambda^0} + q^2}{m_{\Xi'_b}^2 - q^2}\right) 
\biggr] \,, 
\en 
\eq\label{W1LL1}  
W_1^{LL}(q^2) &=& - \sqrt{\dfrac{3}{2}} \, \frac{\beta_{\Lambda^0}}{f_B} \, m_{\neu} \,  
\biggl[ - g^{\tilde bR} \, \dfrac{D+3F}{6} \, 
\dfrac{m_{\Xi_b} + m_{\Lambda^0}}{m_{\Xi_b}^2 - q^2}  
\nonumber\\
&+& g^{\tilde uR} \, \left(
  \dfrac{D+3F}{12} \, \dfrac{m_{\Xi_b} + m_{\Lambda^0}}{m_{\Xi_b}^2 - q^2} 
+ \dfrac{3 (D-F)}{4}  \, \dfrac{m_{\Xi'_b}  + m_{\Lambda^0}}{m_{\Xi'_b}^2 - q^2}\right) 
\nonumber\\
&+& g^{\tilde sR}  \, \left( 
- \dfrac{D+3F}{12} \, \dfrac{m_{\Xi_b} + m_{\Lambda^0}}{m_{\Xi_b}^2 - q^2} 
+ \dfrac{3 (D-F)}{4}  \, \dfrac{m_{\Xi'_b} + m_{\Lambda^0}}{m_{\Xi'_b}^2 - q^2}\right) 
\biggr] \,,
\en 
for the $\Lambda^0$ mode and 
\eq\label{W0LL2}  
W_0^{LL}(q^2) &=& -  \sqrt{\dfrac{1}{2}} \, \frac{\beta_{\Sigma}}{f_B} \, 
\biggl[ g^{\tilde bR} \, \left(1 + \dfrac{D+3F}{3} \, 
\dfrac{m_{\Xi_b} m_{\Sigma^0} + q^2}{m_{\Xi_b}^2 - q^2}\right) 
\nonumber\\
&+& g^{\tilde uR} \, \left(- 1 
- \dfrac{D+3F}{6} \, \dfrac{m_{\Xi_b} m_{\Sigma^0} + q^2}{m_{\Xi_b}^2 - q^2}   
+ \dfrac{D-F}{2}  \, \dfrac{m_{\Xi'_b} m_p + q^2}{m_{\Xi'_b}^2 - q^2}\right) 
\nonumber\\
&+& g^{\tilde sR}  \, \left( 
  \dfrac{D+3F}{6} \, \dfrac{m_{\Xi_b} m_{\Sigma^0} + q^2}{m_{\Xi_b}^2 - q^2}  
+ \dfrac{D-F}{2}   \, \dfrac{m_{\Xi'_b} m_{\Sigma^0} + q^2}{m_{\Xi'_b}^2 - q^2}\right) 
\biggr] \,, 
\en 
\eq\label{W1LL2}  
W_1^{LL}(q^2) &=& - \sqrt{\dfrac{1}{2}} \frac{\beta_{\Sigma}}{f_B} \, m_{\neu} \,  
\biggl[ g^{\tilde bR} \, \dfrac{D+3F}{3} \, 
\dfrac{m_{\Xi_b} + m_{\Sigma^0}}{m_{\Xi_b}^2 - q^2}  
\nonumber\\
&+& g^{\tilde uR} \, \left(
- \dfrac{D+3F}{6} \, \dfrac{m_{\Xi_b} + m_{\Sigma^0}}{m_{\Xi_b}^2 - q^2} 
+ \dfrac{D-F}{2}  \, \dfrac{m_{\Xi'_b}  + m_{\Sigma^0}}{m_{\Xi'_b}^2 - q^2}\right) 
\nonumber\\
&+& g^{\tilde sR}  \, \left( 
  \dfrac{D+3F}{6} \, \dfrac{m_{\Xi_b} + m_{\Sigma^0}}{m_{\Xi_b}^2 - q^2} 
+ \dfrac{D-F}{2}  \, \dfrac{m_{\Xi'_b} + m_{\Sigma^0}}{m_{\Xi'_b}^2 - q^2}\right) 
\biggr] \,,
\en 
for the $\Sigma^0$ mode.
For the masses of the $\Lambda^0$ and $\Sigma^0$, we take the central values from the Particle Data Group~\cite{Zyla:2020zbs}: $m_{\Lambda^0} = 1.115683$ GeV and $m_{\Sigma^0} = 1.192642$ GeV.

\bibliographystyle{JHEP}
\bibliography{main}

\end{document}